\documentclass[twocolumn,aps,prl,10pt,superscriptaddress,longbibliography,floatfix,citeautoscript]{revtex4-2}

\usepackage{amsmath}
\usepackage{amssymb}
\usepackage{bm}
\usepackage{dcolumn}
\usepackage{glossaries}
\usepackage{graphicx}
\usepackage{multirow}
\usepackage{fancyvrb}
\usepackage[linesnumbered,ruled]{algorithm2e}
\usepackage{siunitx}
\usepackage{bibunits}

\usepackage[usenames,dvipsnames,svgnames]{xcolor}
\usepackage{hyperref}
\hypersetup{
    pdfnewwindow=true,      
    colorlinks=true,        
    linkcolor=Blue,         
    citecolor=Blue,         
    filecolor=Blue,         
    urlcolor=Blue           
}

\usepackage{listings}	
\newacronym{bcc}{BCC}{body-centered cubic}
\newacronym{dft}{DFT}{density-functional theory}
\newacronym{dp}{DP}{deep potential}
\newacronym{eam}{EAM}{embedded-atom method}
\newacronym{mc}{MC}{Monte Carlo}
\newacronym{md}{MD}{molecular dynamics}
\newacronym{ml}{ML}{machine learning}
\newacronym{mlp}{MLP}{machine-learned potential}
\newacronym{mpea}{MPEA}{multi-principal-element alloy}
\newacronym{nep}{NEP}{neuroevolution potential}
\newacronym{nn}{NN}{neural network}
\newacronym{pc}{PC}{principal component}
\newacronym{rmse}{RMSE}{root-mean-square error}
\newacronym{snes}{SNES}{separable natural evolution strategy}
\newacronym{sro}{SRO}{short-range order}
\newacronym{apt}{APT}{atom probe tomography}
\newacronym{exafs}{EXAFS}{Extended x-Ray Absorption Fine Structure}

\DeclareSIUnit\angstrom{\text{Å}}
\DeclareSIUnit{\atom}{atom}
\DeclareSIUnit{\step}{step}
\DeclareSIUnit{\atomstepsecond}{\atom\step\per\second}

\newcolumntype{d}{D{.}{.}{-1}}

\begin{document}


\title{Anisotropic Short-Range Order Modulates Ferroelectric Switching in Wurtzite ScAlN Alloys}

\author{Shunda Chen}
\email{shdchen@email.gwu.edu}
\affiliation{Department of Civil and Environmental Engineering, George Washington University, Washington, DC 20052, USA}

\author{Xianchao Dong}
\affiliation{Department of Civil and Environmental Engineering, George Washington University, Washington, DC 20052, USA}

\author{Xiaochen Jin}
\affiliation{Department of Civil and Environmental Engineering, George Washington University, Washington, DC 20052, USA}

\author{Tianshu Li}
\email{tsli@email.gwu.edu}
\affiliation{Department of Civil and Environmental Engineering, George Washington University, Washington, DC 20052, USA}

\date{\today}

\begin{abstract}

Ferroelectric switching in wurtzite alloys is typically understood in terms of composition, strain, defects, and interfaces, while local chemical order is often neglected or treated as a secondary perturbation. Here we show that short-range order (SRO) is a previously overlooked microscopic variable that substantially influences the intrinsic switching barrier. Using first-principles canonical sampling, we find that wurtzite ScAlN develops a robust, highly anisotropic SRO that challenges the conventional random-alloy picture. This ordering suppresses in-plane Sc--N--Sc motifs while enhancing columnar mixed-cation chains along the polar $c$ axis, reflecting the symmetry-distinct polar and basal directions of the wurtzite lattice and reorganizing its polar connectivity. Relative to random-alloy structures, SRO systematically increases the intrinsic switching barrier across a broad composition range. Motif-resolved analysis further identifies the population of columnar Sc--N--Al--N--Sc motifs as the primary structural descriptor underlying switching-barrier variations among configurations with different local order. These results establish anisotropic SRO as an independent degree of freedom for tuning ferroelectric switching. More broadly, they reveal how local chemical order can couple to the symmetry-distinct directions of a polar semiconductor lattice to modify functional behavior. Our findings lay a foundation for SRO engineering as a route to tailoring switching barriers without changing alloy composition.

\end{abstract}

\begin{bibunit}[apsrev4-2]

\maketitle


\emph{Introduction---}Scandium aluminum nitride (ScAlN) alloys have recently emerged as a leading wurtzite ferroelectric semiconductor \cite{Fichtner2019AlScN,Jena2019thenewnitrides,Wang2020FerroelectricSwitching,Mikolajick2021next,Rassay2021ASegmented,wang2021FullyMBE,trolier-mckinstry_new_2025}. Unlike conventional oxide ferroelectrics, ScAlN combines robust ferroelectricity with compatibility with complementary metal–oxide–semiconductor (CMOS) processing, together with excellent thermal and mechanical stability \cite{Islam_APL_2021_on,Guido_ACSAMI_thermal_2023,wang2023dawn}. These attributes make ScAlN attractive for integrated electronics and operation in harsh environments \cite{kim2023wurtziteNatNano,Wang_Mi_APL2024_Perspectives,Casamento_APL_2024_Perspectives,wang2025electric,Li_science_2025_review,Fichtner_2025_APR,cho_write_NC_2026}. 

Although ferroelectric switching and compositionally tunable polarization have been widely demonstrated, the microscopic structural factors governing the intrinsic switching barrier remain poorly understood \cite{Messi_APLM_2025,HIRATA_MT_2025,Behrendt_PRL_2026_Fractals}. In particular, the role of alloy disorder and local bonding topology in determining polarization switching has remained largely unexplored.
Most theoretical descriptions of ScAlN and related wurtzite III-nitride alloys assume a random solid solution \cite{Tasnadi-PRL-2010-origin,zhang2013tunable,Caro_JPCM-2015-Piezoelectric,Talley-2018-PRM-implications,Kyrtsos_2019_PRB_first,Moriwake_2020_APLM_computational,Wang_2021_JAP_Piezoelectric,liu_multiscale_2021,Furuta_2021_JAP_first,Urban_2021_PRB_first,Ye_2021_PSSR_atomistic,Wang_2022_AIPA_understanding,Balestra_2022_JAP_electron,yazawa_JMCC_2022_local,Wang_APL_2024_towards,Lee_sciadv_2024_switching,Pike_PRB_2025_understanding,Wang_PRM_2025_structural,Chen_PRM_2025_towards,Safaltin_Acta_2026_tailoring} and are often modeled using special quasirandom structures (SQS) \cite{Zunger_PRL_1990_SQS}. While computationally convenient, this approximation neglects correlated atomic arrangements and particularly atomic short-range order (SRO). In many alloy systems, deviations from random mixing substantially influence structural, electronic, and functional properties \cite{ji_hidden_NC_2019,Cao2020ACSAMI,Zhang_Natue_2020_Short,Roychowdhury_science_2021_enhanced,Jiang_Science_2021_high,jin2021short,jin2022coexistence,chen2023impacts,Corley-Wiciak2023PRApplied,JinChenLiPRM2023,chen2024intricate,liang2024group,LIANG2025MT,vogl2025identification}. Atomic ordering therefore represents an additional microscopic degree of freedom capable of fundamentally altering materials behavior \cite{Jin2025IEEE,LIANG2025MT,vogl2025identification,liu2026_APT-GeSn-CVD-MBE,Anis2026shining}.
Recent experimental and theoretical studies have begun to explore nonrandom Sc distributions in ScAlN, including Sc-rich clustering \cite{bhattarai_effect_2024,hart2025enhanced,Das2026-APL_APT_ScAlN} and growth-dependent chemical inhomogeneity \cite{hart2025enhanced}, with potential consequences for dielectric, electronic, and ferroelectric properties. These observations suggest that the random-alloy approximation may not fully capture the local structure of ScAlN. A fundamental unresolved question is how local chemical order develops within the intrinsically polar wurtzite lattice and whether such ordering acts as an independent control parameter governing ferroelectric switching. Addressing this question is essential for developing a predictive understanding of ferroelectric switching in ScAlN and related polar semiconductors.


\begin{figure*}
\begin{center}
\includegraphics[width=0.90\linewidth]{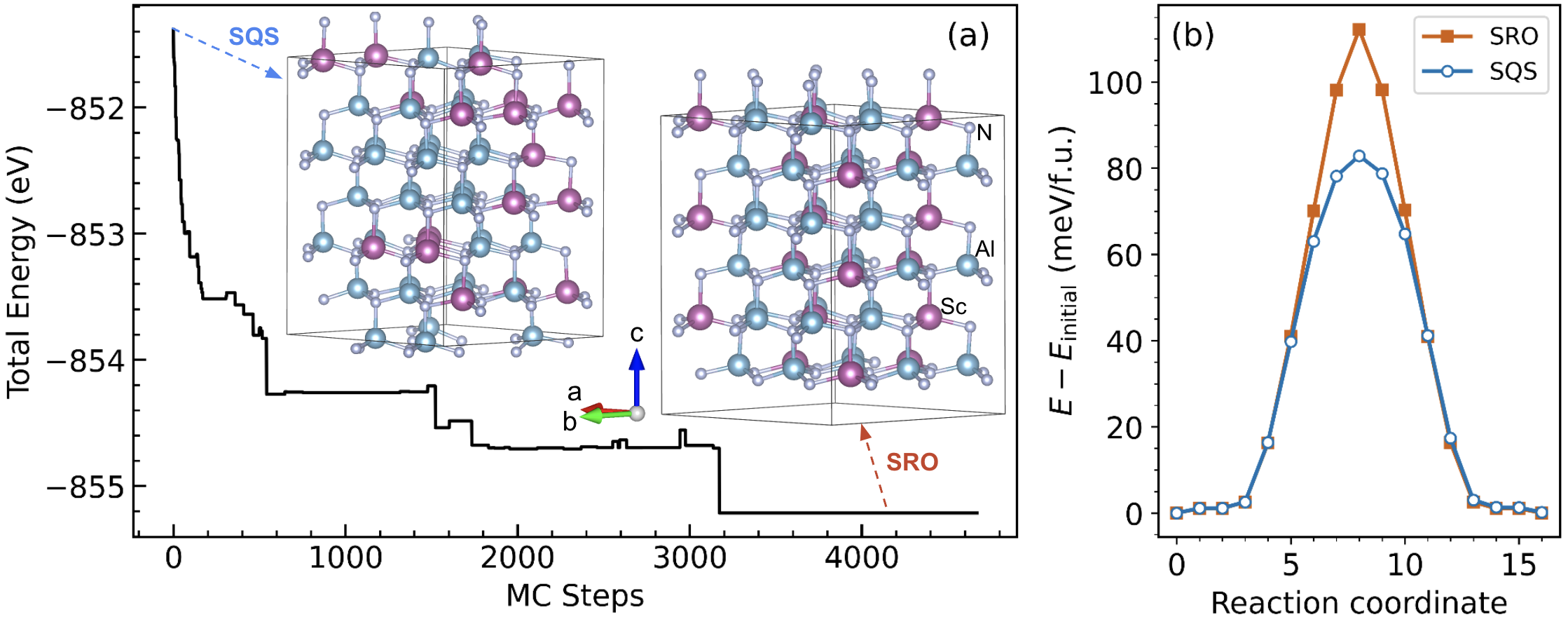}

\caption{\label{MCDFT-sampling} \textbf{Emergence of short-range order and its impact on ferroelectric switching in ScAlN alloys.}
(a) Evolution of the relative total energy during DFT Monte Carlo sampling at 300 K for Sc$_{0.33}$Al$_{0.67}$N, starting from an initial special quasirandom structure (SQS) configuration and evolving toward an energetically favorable short-range-order (SRO) state. Insets show representative atomic configurations of the initial SQS and an SRO structure, highlighting the emergence of columnar ordering along the $c$ axis in the SRO configuration. (b) Minimum-energy switching pathways obtained from solid-state nudged elastic band (SS-NEB) calculations for representative SQS and SRO structures at $x=0.33$. The SRO structure exhibits a substantially larger switching barrier, demonstrating that SRO strongly influences ferroelectric switching barriers.
}

\end{center}
\end{figure*} 


\begin{figure}
\begin{center}
\includegraphics[width=1.0\linewidth]{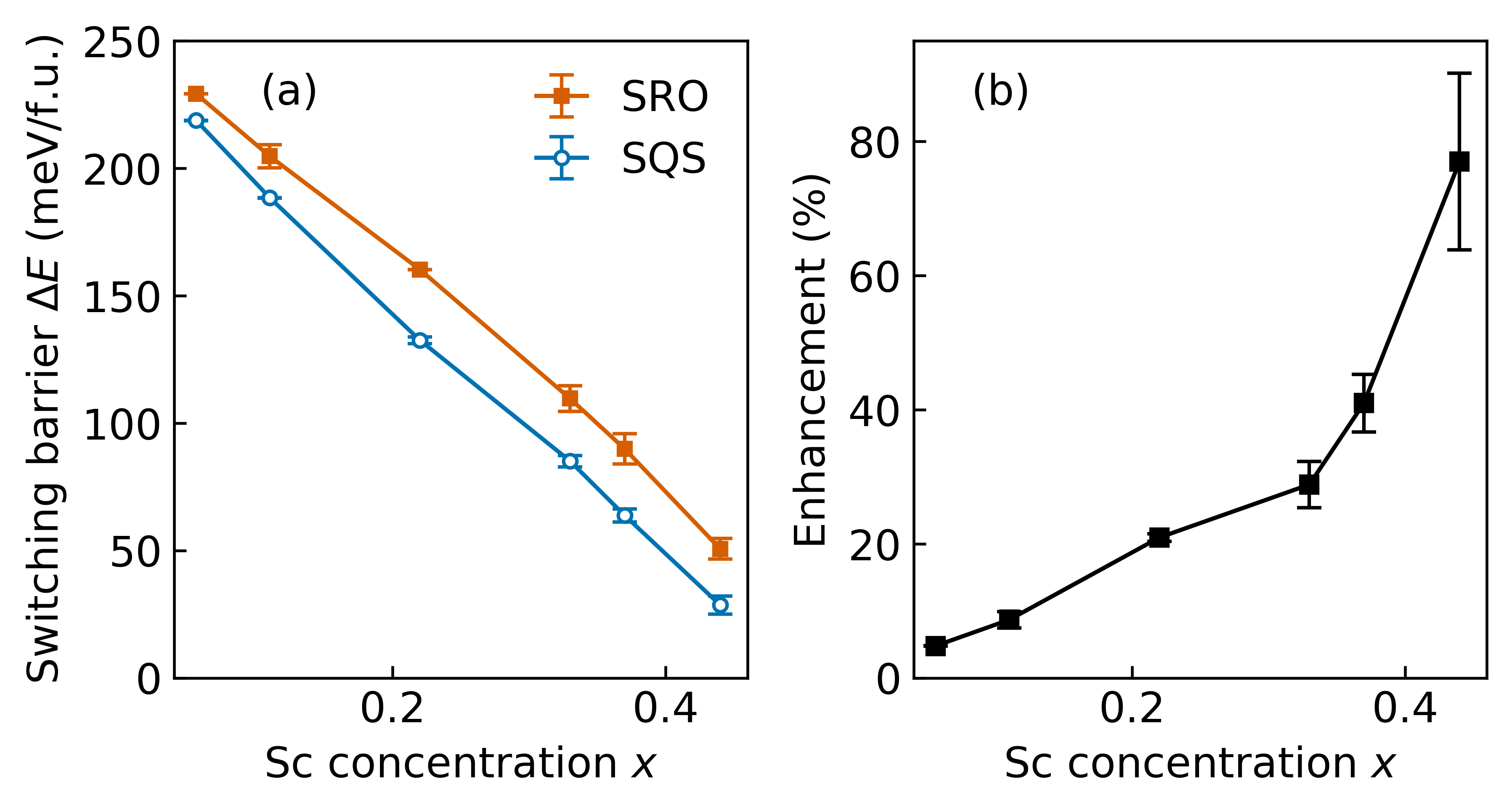}

\caption{\label{switching-barrier-x} \textbf{Composition dependence of ferroelectric switching barriers in Sc$_x$Al$_{1-x}$N alloys.}
(a) Ferroelectric switching barriers as a function of Sc composition for special quasirandom structures (SQS) and short-range-order (SRO) structures. While the switching barrier generally decreases with increasing Sc concentration, SRO systematically increases the barrier relative to random-alloy configurations. Symbols represent averages over multiple independent structures, and error bars denote the standard deviation. (b) Relative enhancement of the switching barrier induced by SRO, defined as $100\%(\Delta E_{\mathrm{SRO}}/\Delta E_{\mathrm{SQS}}-1)$. Error bars denote the propagated standard error of the mean. The influence of SRO becomes increasingly pronounced at intermediate and high Sc concentrations, highlighting the growing importance of local chemical order in high-Sc-content ScAlN.
}

\end{center}
\end{figure} 

Here, we address this fundamental question by combining extensive first-principles canonical sampling and solid-state nudged elastic band calculations. We find that wurtzite ScAlN alloys exhibit a robust and highly anisotropic form of SRO. Contrary to the conventional random-alloy assumption, SRO suppresses in-plane Sc–N–Sc motifs while simultaneously enhancing columnar mixed-cation motifs along the polar $c$ axis. This anisotropic motif network emerges as a robust structural signature of ordering across the investigated composition range and systematically increases ferroelectric switching barriers relative to random-alloy structures. We further identify the population of columnar Sc–N–Al–N–Sc motifs as the primary descriptor of the switching barrier across distinct alloy configurations, establishing a direct link between anisotropic SRO and ferroelectric switching energetics. These results establish anisotropic SRO as a previously overlooked degree of freedom modulating ferroelectric switching. More broadly, this study provides a predictive framework for understanding ferroelectric switching in wurtzite ScAlN and related polar semiconductors, and suggests SRO engineering as a pathway for tuning functional properties without changing alloy composition.

\begin{figure*}
\begin{center}
\includegraphics[width=0.96\linewidth]{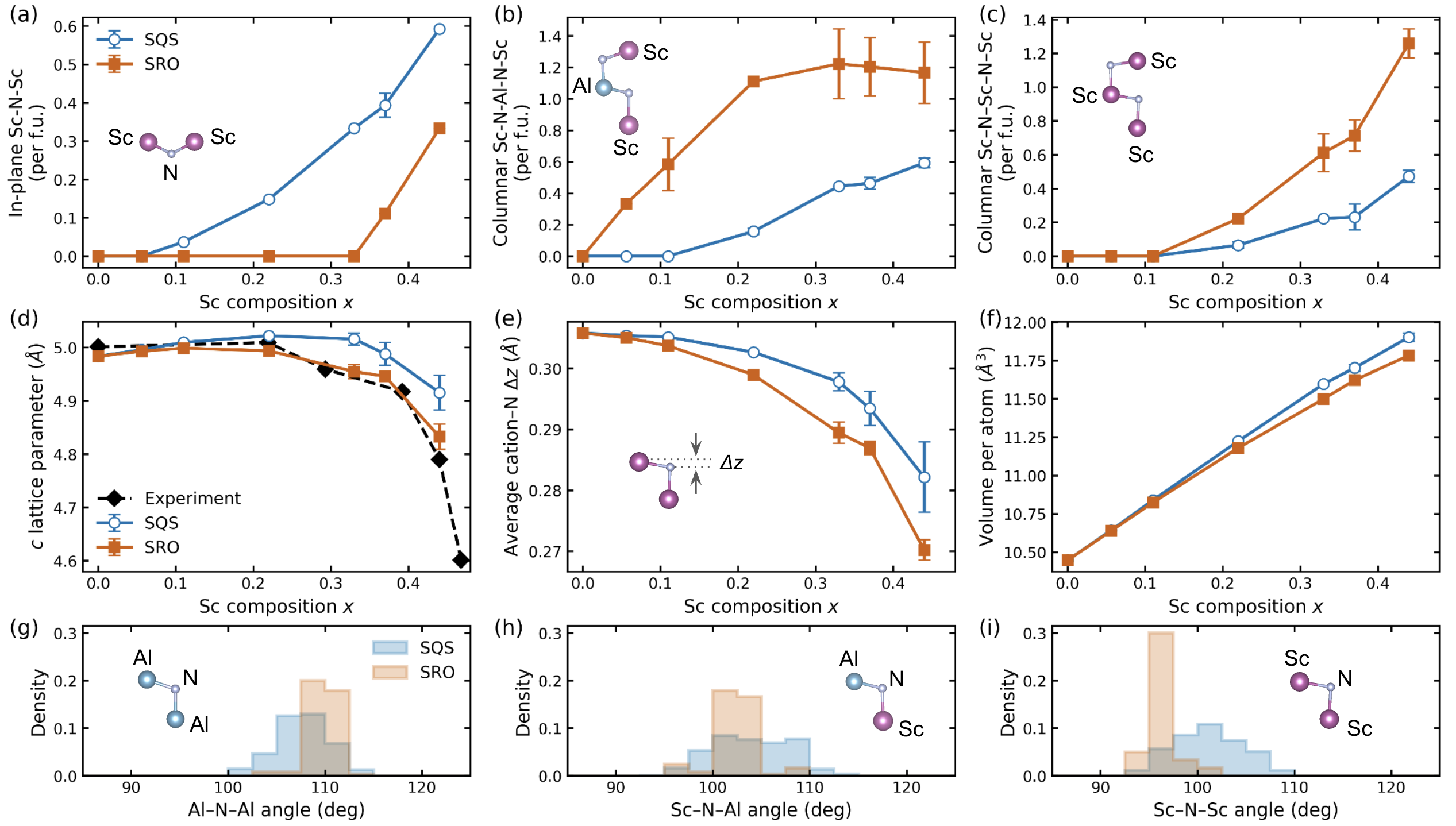}
\caption{\label{mechanisms} \textbf{Anisotropic short-range order reorganizes motif populations and structural properties}.
(a) In-plane Sc--N--Sc motifs per formula unit. (b) Columnar Sc--N--Al--N--Sc motifs per formula unit. (c) Columnar Sc--N--Sc--N--Sc motifs per formula unit. (d) $c$-lattice parameter as a function of Sc composition; experimental values from Ref.~\onlinecite{Akiyama2009AMenhancement} are shown for comparison. (e) Average cation--N vertical displacement $\Delta z$. (f) Volume per atom. Error bars denote the standard deviation across sampled configurations. (g)--(i) Normalized bond-angle distributions for cross-plane Al--N--Al, Sc--N--Al, and Sc--N--Sc bond angles (see insets for illustrative graphs), respectively, obtained from representative structures at $x=0.33$. Relative to SQS alloys, anisotropic SRO suppresses in-plane Sc--N--Sc motifs while promoting columnar Sc--N--Al--N--Sc and Sc--N--Sc--N--Sc motifs. This motif reorganization is accompanied by contraction of the $c$ lattice parameter, reduced cation--N vertical displacement, smaller volume per atom, and narrower cross-plane metal--N--metal bond-angle distributions.}
\end{center}
\end{figure*} 

%

\emph{Emergence of short-range order---}
To determine whether ScAlN alloys can be approximated as a random solid solution, we investigated atomic configurations using first-principles canonical ensemble sampling based on DFT Monte Carlo (MC/DFT) simulations (see SI for methods). Starting from initial SQS configurations, MC/DFT sampling consistently drives the system toward lower-energy SRO states.

As shown in Fig.~\ref{MCDFT-sampling}(a), the total energy decreases substantially during MC/DFT sampling for Sc$_{0.33}$Al$_{0.67}$N at 300 K, indicating that correlated atomic arrangements are energetically favored over random-alloy configurations. Importantly, the SRO structures discussed here are not imposed \emph{a priori} but emerge spontaneously during canonical sampling while lowering the total energy relative to the initial SQS configurations, demonstrating that they are thermodynamically favored local arrangements. The resulting SRO structures exhibit pronounced local structural correlations and anisotropic ordering along the wurtzite $c$ axis. Representative snapshot in Fig.~\ref{MCDFT-sampling}(a) reveals the emergence of columnar Sc-containing arrangements extending along the polar direction, in contrast to the SQS structure.

The emergence of SRO has immediate implications for ferroelectric switching. Figure~\ref{MCDFT-sampling}(b) compares representative minimum-energy switching pathways obtained from solid-state nudged elastic band (SS-NEB) calculations for SQS and SRO structures at $x=0.33$ (see SI for methods and supplementary Fig.~\ref{fig-convergence} for convergence tests). Although both structures switch through the same wurtzite polarization-reversal pathway, the SRO configuration exhibits a substantially larger switching barrier. This result demonstrates that SRO directly modifies the ferroelectric energy landscape and suggests that the random-alloy approximation fails to accurately capture key switching physics in ScAlN.


\emph{Composition dependence of switching barriers---}
To establish the generality of the SRO effect, we systematically computed ferroelectric switching barriers across a broad composition range using multiple independent SQS and SRO configurations for each composition (see, for example, supplementary Fig.~\ref{fig-mc-Sc0.33} and Fig.~\ref{fig-mc-Sc0.37}). As shown in Fig.~\ref{switching-barrier-x}(a), the switching barrier decreases overall with increasing Sc concentration, consistent with the progressive destabilization of the polar wurtzite phase at high Sc content. However, at every composition examined, SRO systematically increases the switching barrier relative to the corresponding SQS structures.

Figure~\ref{switching-barrier-x}(b) further highlights the relative enhancement induced by SRO. Depending on composition, SRO increases the switching barrier by up to about 80\% relative to random-alloy predictions. The enhancement becomes increasingly pronounced at intermediate and high Sc concentrations, demonstrating that atomic ordering is not a minor perturbation but a major factor affecting ferroelectric switching energetics. 




\begin{figure*}
\begin{center}

\includegraphics[width=0.95\linewidth]{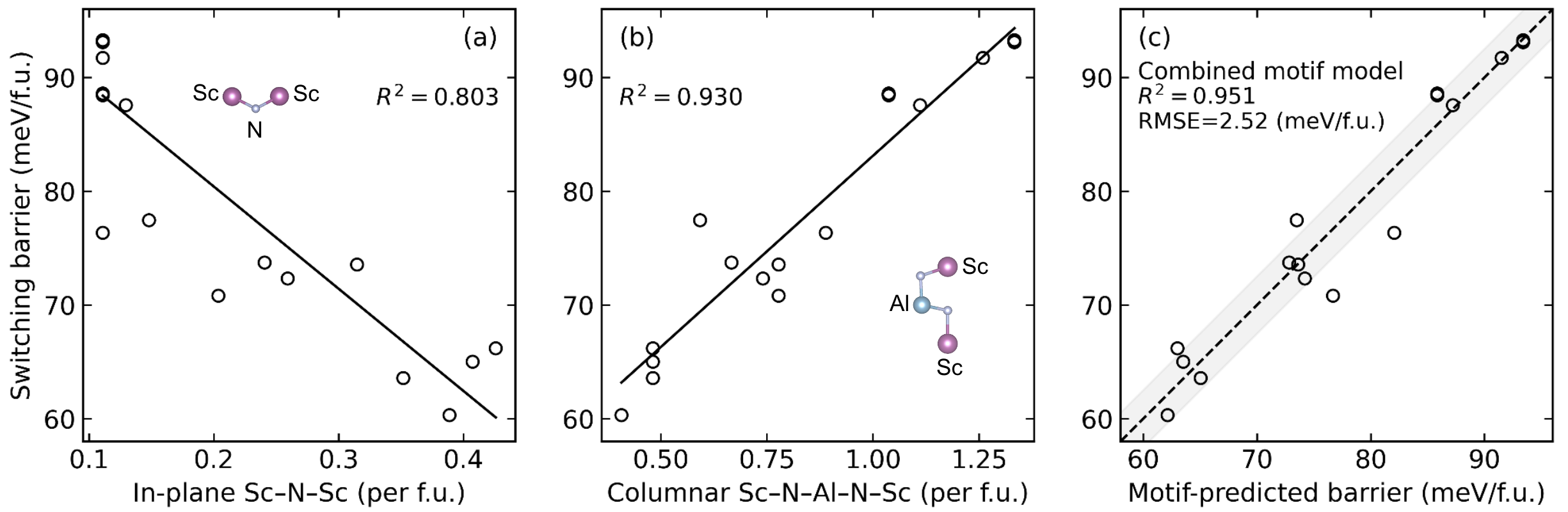}

\caption{\label{fig4}
\textbf{Anisotropic short-range order tunes switching barriers at fixed composition.}
Switching barrier from SS-NEB calculations as a function of the population of (a) in-plane Sc--N--Sc motifs per formula unit (f.u.) and (b) columnar Sc--N--Al--N--Sc motifs per f.u. Solid lines denote linear least-squares fits. The columnar Sc--N--Al--N--Sc motif exhibits the strongest correlation with the switching barrier ($R^2=0.930$). (c) Parity plot comparing SS-NEB switching barriers with predictions from a linear model using the populations of in-plane Sc--N--Sc and columnar Sc--N--Al--N--Sc motifs as descriptors. The shaded region denotes the root-mean-square error (RMSE). All data correspond to Sc$_{0.37}$Al$_{0.63}$N configurations with varying degree of anisotropic SRO. The combined motif model reproduces the switching barriers with high accuracy ($R^2=0.951$), demonstrating that anisotropic SRO serves as an independent structural degree of freedom for tuning ferroelectric switching barriers.
}

\end{center}
\end{figure*} 


\emph{Anisotropic short-range order tunes switching barriers---}
To uncover the microscopic origin of the enhanced switching barriers, we analyzed characteristic local motifs and structural properties. Figure~\ref{mechanisms}(a) shows that SRO strongly suppresses the population of in-plane Sc--N--Sc motifs relative to SQS structures, particularly at low and intermediate Sc concentrations. In contrast, SRO promotes columnar Sc-containing motifs extending along the polar $c$ axis, including Sc--N--Al--N--Sc motifs [Fig.~\ref{mechanisms}(b)] and Sc--N--Sc--N--Sc motifs [Fig.~\ref{mechanisms}(c)]. These trends reveal that SRO is highly anisotropic, suppressing in-plane Sc connectivity while enhancing polar connectivity along the crystallographic $c$ direction. 

The energetic origin of this anisotropic ordering can be elucidated through a controlled motif comparison. We constructed three otherwise identical 108-atom Sc$_{0.037}$Al$_{0.963}$N supercells differing only in the placement of a single local motif. After full structural relaxation, relative to an in-plane Sc--N--Sc motif, a cross-plane Sc--N--Sc motif lowers the total energy by 85.2 meV (see supplementary Fig.~\ref{fig-motif-cross-Sc-N-Sc}), while a columnar Sc--N--Al--N--Sc motif lowers the total energy by 214.5 meV (see supplementary Fig.~\ref{fig-motif-columnar-Sc-N-Al-N-Sc}). These results reveal a clear energetic hierarchy in which laterally connected Sc-rich motifs are least favorable, cross-plane Sc connectivity is intermediate, and columnar mixed-cation connectivity is most stable. Because the polar wurtzite lattice (space group $P6_3mc$) distinguishes the basal plane from the polar [0001] direction, local chemical correlations parallel and perpendicular to the $c$ axis belong to  symmetry-inequivalent classes. The motif hierarchy demonstrates that these symmetry-distinct correlation channels possess markedly different energetic preferences. Anisotropic SRO can therefore be rationalized as an energetic response that suppresses laterally connected Sc environments and promotes energetically favorable columnar motifs along the polar direction. The results are consistent with reduced local bonding frustration associated with accommodating Sc-rich environments within the polar wurtzite lattice, providing a microscopic explanation for the anisotropic SRO observed in the MC sampling.

This motif reorganization is accompanied by systematic changes in alloy structures. Relative to SQS alloys, SRO produces a more contracted $c$ lattice parameter [Fig.~\ref{mechanisms}(d)], reduced cation--N vertical displacement $\Delta z$ [Fig.~\ref{mechanisms}(e)], a smaller volume per atom [Fig.~\ref{mechanisms}(f)], and narrower cross-plane metal--N--metal bond-angle distributions [Fig.~\ref{mechanisms}(g)--(i)] (see supplementary Fig.~\ref{fig-angle} for the bond-angle distributions across compositions). Notably, the SRO-derived $c$ lattice parameters agree substantially better with experiment than the SQS results, providing independent evidence that the SRO atomic arrangements are physically relevant in ScAlN. Together, these results indicate that anisotropic SRO suppresses local structural disorder and reorganizes the polar wurtzite framework into a more compact and directionally correlated network, whose local chemical correlations reflect the symmetry-distinct polar axis and basal plane of the lattice.

To directly connect anisotropic SRO with switching energetics, we examined a series of Sc$_{0.37}$Al$_{0.63}$N configurations with different degrees of local ordering generated through MC/DFT sampling at different temperatures (see SI for methods). As shown in Fig.~\ref{fig4}(a), the switching barrier decreases with increasing in-plane Sc--N--Sc population ($R^2=0.803$). Notably, the switching barrier exhibits a strong positive correlation with the population of columnar Sc--N--Al--N--Sc motifs [Fig.~\ref{fig4}(b)], with $R^2=0.930$. These results identify columnar Sc--N--Al--N--Sc motifs as the primary microscopic structural descriptor linking anisotropic SRO to ferroelectric switching barriers. 

We further constructed a linear model using only the populations of in-plane Sc--N--Sc and columnar Sc--N--Al--N--Sc motifs. The resulting combined motif model accurately reproduces the DFT switching barriers across distinct alloy configurations with $R^2=0.951$ [Fig.~\ref{fig4}(c)]. The success of this combined motif model demonstrates that a small set of local motif populations captures the dominant structural contributions to the intrinsic switching barrier. More broadly, these results establish local anisotropic SRO as an independent degree of freedom for tuning ferroelectric switching barriers at fixed composition. They further demonstrate how local chemical order can couple to the symmetry-distinct directions of a polar lattice to modify functional behavior, providing a microscopic foundation for SRO engineering.


\emph{Conclusion and discussion---}In summary, we demonstrate that wurtzite ScAlN exhibits a highly anisotropic SRO which substantially modifies the intrinsic ferroelectric switching barrier. SRO suppresses in-plane Sc-N-Sc motifs while enhancing columnar mixed-cation motifs along the polar c axis. This anisotropic motif reorganization contracts the lattice, narrows cross-plane bond-angle distributions, and produces structural parameters that are closer to experiment than those obtained from random-alloy configurations.

The physical origin of this ordering can be rationalized in terms of the structural incompatibility between AlN-like tetrahedral bonding and Sc-rich local bonding preferences within the wurtzite framework. AlN strongly stabilizes a polar tetrahedral network, whereas Sc incorporation introduces local bonding environments that are less compatible with the ideal wurtzite geometry. Because the polar wurtzite lattice distinguishes the basal plane from the unique [0001] direction, local chemical correlations parallel and perpendicular to the polar axis are symmetry inequivalent, allowing these energetic preferences to manifest as anisotropic chemical ordering. Anisotropic SRO provides a local strain-relief pathway by avoiding laterally connected Sc-rich environments and promoting columnar mixed-metal connectivity along the polar direction. From this perspective, the preferred SRO motifs can be viewed as structural responses to local bonding frustration in the polar lattice.

This motif-level view further explains the strong dependence of the ferroelectric switching barrier on chemical order. Polarization reversal in wurtzite ferroelectrics involves substantial distortions of the metal-nitrogen framework, whose local connectivity is reorganized by SRO. As a result, local ordering patterns reshape the energy landscape along the switching pathway. In particular, columnar mixed-metal connectivity along the polar direction appears to stabilize the polar network against reversal, leading to a higher intrinsic switching barrier. Consistent with this interpretation, we find that, at fixed composition, the barrier is most strongly correlated with the population of columnar Sc--N--Al--N--Sc motifs. Furthermore, a simple two-motif model accurately reproduces the first-principles barriers, demonstrating that a small set of local motif populations captures the dominant structural contributions to the intrinsic switching barrier.

Recent characterization studies have begun to reveal nanoscale, nonrandom Sc distributions in ScAlN \cite{hart2025enhanced,Das2026-APL_APT_ScAlN}, providing experimental evidence that local chemical order is a physically relevant feature of this alloy. The present work offers a microscopic framework for interpreting such observations by both identifying the anisotropic motifs and quantifying their impact on intrinsic switching barriers. Recent atomistic studies indicate that ferroelectric switching in ScAlN proceeds through highly localized structural transformations \cite{Huang_PRL_2026_atomistic,Zheng_PRL_2026_Domain}, and our results suggest that local SRO should therefore be considered as an essential structural variable in understanding ferroelectricity in wurtzite alloys. 

More broadly, our findings establish anisotropic SRO as a previously overlooked structural degree of freedom for engineering ferroelectric behavior. Because local chemical order can in principle be influenced by growth conditions \cite{liu2026_APT-GeSn-CVD-MBE}, annealing \cite{Anis2026shining}, and other processing pathways \cite{Vogl_MM_2024_exploring}, SRO engineering may provide a new route to tailoring ferroelectricity, alongside established design variables such as composition and strain. At a fundamental level, our results demonstrate how local chemical order can couple to the symmetry-distinct directions of a polar lattice to modify functional behavior. Because this mechanism originates from the interplay between local chemical ordering and crystallographic anisotropy, analogous anisotropic ordering phenomena may arise in a broad range of polar semiconductor and ferroelectric alloys. Extending this concept to other alloy systems could enable materials-by-design strategies in which composition, strain, defects, interfaces, and local chemical order act as coupled design variables.


\vspace{1em}


\emph{Acknowledgments---} The authors thank Zetian Mi, Nicholas Barrett, Hongping Zhao, 
and Deep Jariwala for useful discussions. This work is supported by the Air Force Office of Scientific Research under award number FA9550-25-1-0109.
The authors acknowledge Department of Defense High Performance Computing Modernization Program and GW High Performance Computing for the computing support. 

\vspace{1em}

\emph{Data availability---}The data that support the findings of this study are available from the corresponding author upon reasonable request. 

\putbib[Refs-SRO]

\end{bibunit}


\newpage







\begin{thebibliography}{65}%
\makeatletter
\providecommand \@ifxundefined [1]{%
 \@ifx{#1\undefined}
}%
\providecommand \@ifnum [1]{%
 \ifnum #1\expandafter \@firstoftwo
 \else \expandafter \@secondoftwo
 \fi
}%
\providecommand \@ifx [1]{%
 \ifx #1\expandafter \@firstoftwo
 \else \expandafter \@secondoftwo
 \fi
}%
\providecommand \natexlab [1]{#1}%
\providecommand \enquote  [1]{``#1''}%
\providecommand \bibnamefont  [1]{#1}%
\providecommand \bibfnamefont [1]{#1}%
\providecommand \citenamefont [1]{#1}%
\providecommand \href@noop [0]{\@secondoftwo}%
\providecommand \href [0]{\begingroup \@sanitize@url \@href}%
\providecommand \@href[1]{\@@startlink{#1}\@@href}%
\providecommand \@@href[1]{\endgroup#1\@@endlink}%
\providecommand \@sanitize@url [0]{\catcode `\\12\catcode `\$12\catcode `\&12\catcode `\#12\catcode `\^12\catcode `\_12\catcode `\%12\relax}%
\providecommand \@@startlink[1]{}%
\providecommand \@@endlink[0]{}%
\providecommand \url  [0]{\begingroup\@sanitize@url \@url }%
\providecommand \@url [1]{\endgroup\@href {#1}{\urlprefix }}%
\providecommand \urlprefix  [0]{URL }%
\providecommand \Eprint [0]{\href }%
\providecommand \doibase [0]{https://doi.org/}%
\providecommand \selectlanguage [0]{\@gobble}%
\providecommand \bibinfo  [0]{\@secondoftwo}%
\providecommand \bibfield  [0]{\@secondoftwo}%
\providecommand \translation [1]{[#1]}%
\providecommand \BibitemOpen [0]{}%
\providecommand \bibitemStop [0]{}%
\providecommand \bibitemNoStop [0]{.\EOS\space}%
\providecommand \EOS [0]{\spacefactor3000\relax}%
\providecommand \BibitemShut  [1]{\csname bibitem#1\endcsname}%
\let\auto@bib@innerbib\@empty
\bibitem [{\citenamefont {Fichtner}\ \emph {et~al.}(2019)\citenamefont {Fichtner}, \citenamefont {Wolff}, \citenamefont {Lofink}, \citenamefont {Kienle},\ and\ \citenamefont {Wagner}}]{Fichtner2019AlScN}%
  \BibitemOpen
  \bibfield  {author} {\bibinfo {author} {\bibfnamefont {S.}~\bibnamefont {Fichtner}}, \bibinfo {author} {\bibfnamefont {N.}~\bibnamefont {Wolff}}, \bibinfo {author} {\bibfnamefont {F.}~\bibnamefont {Lofink}}, \bibinfo {author} {\bibfnamefont {L.}~\bibnamefont {Kienle}},\ and\ \bibinfo {author} {\bibfnamefont {B.}~\bibnamefont {Wagner}},\ }\href {https://doi.org/10.1063/1.5084945} {\bibfield  {journal} {\bibinfo  {journal} {Journal of Applied Physics}\ }\textbf {\bibinfo {volume} {125}},\ \bibinfo {pages} {114103} (\bibinfo {year} {2019})}\BibitemShut {NoStop}%
\bibitem [{\citenamefont {Jena}\ \emph {et~al.}(2019)\citenamefont {Jena}, \citenamefont {Page}, \citenamefont {Casamento}, \citenamefont {Dang}, \citenamefont {Singhal}, \citenamefont {Zhang}, \citenamefont {Wright}, \citenamefont {Khalsa}, \citenamefont {Cho},\ and\ \citenamefont {Xing}}]{Jena2019thenewnitrides}%
  \BibitemOpen
  \bibfield  {author} {\bibinfo {author} {\bibfnamefont {D.}~\bibnamefont {Jena}}, \bibinfo {author} {\bibfnamefont {R.}~\bibnamefont {Page}}, \bibinfo {author} {\bibfnamefont {J.}~\bibnamefont {Casamento}}, \bibinfo {author} {\bibfnamefont {P.}~\bibnamefont {Dang}}, \bibinfo {author} {\bibfnamefont {J.}~\bibnamefont {Singhal}}, \bibinfo {author} {\bibfnamefont {Z.}~\bibnamefont {Zhang}}, \bibinfo {author} {\bibfnamefont {J.}~\bibnamefont {Wright}}, \bibinfo {author} {\bibfnamefont {G.}~\bibnamefont {Khalsa}}, \bibinfo {author} {\bibfnamefont {Y.}~\bibnamefont {Cho}},\ and\ \bibinfo {author} {\bibfnamefont {H.~G.}\ \bibnamefont {Xing}},\ }\href {https://doi.org/10.7567/1347-4065/ab147b} {\bibfield  {journal} {\bibinfo  {journal} {Japanese Journal of Applied Physics}\ }\textbf {\bibinfo {volume} {58}},\ \bibinfo {pages} {SC0801} (\bibinfo {year} {2019})}\BibitemShut {NoStop}%
\bibitem [{\citenamefont {Wang}\ \emph {et~al.}(2020)\citenamefont {Wang}, \citenamefont {Zheng}, \citenamefont {Musavigharavi}, \citenamefont {Zhu}, \citenamefont {Foucher}, \citenamefont {Trolier-McKinstry}, \citenamefont {Stach},\ and\ \citenamefont {Olsson}}]{Wang2020FerroelectricSwitching}%
  \BibitemOpen
  \bibfield  {author} {\bibinfo {author} {\bibfnamefont {D.}~\bibnamefont {Wang}}, \bibinfo {author} {\bibfnamefont {J.}~\bibnamefont {Zheng}}, \bibinfo {author} {\bibfnamefont {P.}~\bibnamefont {Musavigharavi}}, \bibinfo {author} {\bibfnamefont {W.}~\bibnamefont {Zhu}}, \bibinfo {author} {\bibfnamefont {A.~C.}\ \bibnamefont {Foucher}}, \bibinfo {author} {\bibfnamefont {S.~E.}\ \bibnamefont {Trolier-McKinstry}}, \bibinfo {author} {\bibfnamefont {E.~A.}\ \bibnamefont {Stach}},\ and\ \bibinfo {author} {\bibfnamefont {R.~H.}\ \bibnamefont {Olsson}},\ }\href {https://doi.org/10.1109/LED.2020.3034576} {\bibfield  {journal} {\bibinfo  {journal} {IEEE Electron Device Letters}\ }\textbf {\bibinfo {volume} {41}},\ \bibinfo {pages} {1774} (\bibinfo {year} {2020})}\BibitemShut {NoStop}%
\bibitem [{\citenamefont {Mikolajick}\ \emph {et~al.}(2021)\citenamefont {Mikolajick}, \citenamefont {Slesazeck}, \citenamefont {Mulaosmanovic}, \citenamefont {Park}, \citenamefont {Fichtner}, \citenamefont {Lomenzo}, \citenamefont {Hoffmann},\ and\ \citenamefont {Schroeder}}]{Mikolajick2021next}%
  \BibitemOpen
  \bibfield  {author} {\bibinfo {author} {\bibfnamefont {T.}~\bibnamefont {Mikolajick}}, \bibinfo {author} {\bibfnamefont {S.}~\bibnamefont {Slesazeck}}, \bibinfo {author} {\bibfnamefont {H.}~\bibnamefont {Mulaosmanovic}}, \bibinfo {author} {\bibfnamefont {M.~H.}\ \bibnamefont {Park}}, \bibinfo {author} {\bibfnamefont {S.}~\bibnamefont {Fichtner}}, \bibinfo {author} {\bibfnamefont {P.~D.}\ \bibnamefont {Lomenzo}}, \bibinfo {author} {\bibfnamefont {M.}~\bibnamefont {Hoffmann}},\ and\ \bibinfo {author} {\bibfnamefont {U.}~\bibnamefont {Schroeder}},\ }\href {https://doi.org/10.1063/5.0037617} {\bibfield  {journal} {\bibinfo  {journal} {Journal of Applied Physics}\ }\textbf {\bibinfo {volume} {129}},\ \bibinfo {pages} {100901} (\bibinfo {year} {2021})}\BibitemShut {NoStop}%
\bibitem [{\citenamefont {Rassay}\ \emph {et~al.}(2021)\citenamefont {Rassay}, \citenamefont {Hakim}, \citenamefont {Li}, \citenamefont {Forgey}, \citenamefont {Choudhary},\ and\ \citenamefont {Tabrizian}}]{Rassay2021ASegmented}%
  \BibitemOpen
  \bibfield  {author} {\bibinfo {author} {\bibfnamefont {S.}~\bibnamefont {Rassay}}, \bibinfo {author} {\bibfnamefont {F.}~\bibnamefont {Hakim}}, \bibinfo {author} {\bibfnamefont {C.}~\bibnamefont {Li}}, \bibinfo {author} {\bibfnamefont {C.}~\bibnamefont {Forgey}}, \bibinfo {author} {\bibfnamefont {N.}~\bibnamefont {Choudhary}},\ and\ \bibinfo {author} {\bibfnamefont {R.}~\bibnamefont {Tabrizian}},\ }\href {https://doi.org/https://doi.org/10.1002/pssr.202100087} {\bibfield  {journal} {\bibinfo  {journal} {physica status solidi (RRL) – Rapid Research Letters}\ }\textbf {\bibinfo {volume} {15}},\ \bibinfo {pages} {2100087} (\bibinfo {year} {2021})}\BibitemShut {NoStop}%
\bibitem [{\citenamefont {Wang}\ \emph {et~al.}(2021{\natexlab{a}})\citenamefont {Wang}, \citenamefont {Wang}, \citenamefont {Vu}, \citenamefont {Chiang}, \citenamefont {Heron},\ and\ \citenamefont {Mi}}]{wang2021FullyMBE}%
  \BibitemOpen
  \bibfield  {author} {\bibinfo {author} {\bibfnamefont {P.}~\bibnamefont {Wang}}, \bibinfo {author} {\bibfnamefont {D.}~\bibnamefont {Wang}}, \bibinfo {author} {\bibfnamefont {N.~M.}\ \bibnamefont {Vu}}, \bibinfo {author} {\bibfnamefont {T.}~\bibnamefont {Chiang}}, \bibinfo {author} {\bibfnamefont {J.~T.}\ \bibnamefont {Heron}},\ and\ \bibinfo {author} {\bibfnamefont {Z.}~\bibnamefont {Mi}},\ }\href {https://doi.org/10.1063/5.0054539} {\bibfield  {journal} {\bibinfo  {journal} {Applied Physics Letters}\ }\textbf {\bibinfo {volume} {118}},\ \bibinfo {pages} {223504} (\bibinfo {year} {2021}{\natexlab{a}})}\BibitemShut {NoStop}%
\bibitem [{\citenamefont {Trolier-McKinstry}\ \emph {et~al.}(2025)\citenamefont {Trolier-McKinstry}, \citenamefont {Uchida},\ and\ \citenamefont {Yamada}}]{trolier-mckinstry_new_2025}%
  \BibitemOpen
  \bibfield  {author} {\bibinfo {author} {\bibfnamefont {S.}~\bibnamefont {Trolier-McKinstry}}, \bibinfo {author} {\bibfnamefont {K.}~\bibnamefont {Uchida}},\ and\ \bibinfo {author} {\bibfnamefont {T.}~\bibnamefont {Yamada}},\ }\href {https://doi.org/10.1557/s43577-025-00969-w} {\bibfield  {journal} {\bibinfo  {journal} {MRS Bulletin}\ }\textbf {\bibinfo {volume} {50}},\ \bibinfo {pages} {1023} (\bibinfo {year} {2025})}\BibitemShut {NoStop}%
\bibitem [{\citenamefont {Islam}\ \emph {et~al.}(2021)\citenamefont {Islam}, \citenamefont {Wolff}, \citenamefont {Yassine}, \citenamefont {Schönweger}, \citenamefont {Christian}, \citenamefont {Kohlstedt}, \citenamefont {Ambacher}, \citenamefont {Lofink}, \citenamefont {Kienle},\ and\ \citenamefont {Fichtner}}]{Islam_APL_2021_on}%
  \BibitemOpen
  \bibfield  {author} {\bibinfo {author} {\bibfnamefont {M.~R.}\ \bibnamefont {Islam}}, \bibinfo {author} {\bibfnamefont {N.}~\bibnamefont {Wolff}}, \bibinfo {author} {\bibfnamefont {M.}~\bibnamefont {Yassine}}, \bibinfo {author} {\bibfnamefont {G.}~\bibnamefont {Schönweger}}, \bibinfo {author} {\bibfnamefont {B.}~\bibnamefont {Christian}}, \bibinfo {author} {\bibfnamefont {H.}~\bibnamefont {Kohlstedt}}, \bibinfo {author} {\bibfnamefont {O.}~\bibnamefont {Ambacher}}, \bibinfo {author} {\bibfnamefont {F.}~\bibnamefont {Lofink}}, \bibinfo {author} {\bibfnamefont {L.}~\bibnamefont {Kienle}},\ and\ \bibinfo {author} {\bibfnamefont {S.}~\bibnamefont {Fichtner}},\ }\href {https://doi.org/10.1063/5.0053649} {\bibfield  {journal} {\bibinfo  {journal} {Applied Physics Letters}\ }\textbf {\bibinfo {volume} {118}},\ \bibinfo {pages} {232905} (\bibinfo {year} {2021})}\BibitemShut {NoStop}%
\bibitem [{\citenamefont {Guido}\ \emph {et~al.}(2023)\citenamefont {Guido}, \citenamefont {Lomenzo}, \citenamefont {Islam}, \citenamefont {Wolff}, \citenamefont {Gremmel}, \citenamefont {Sch{\"o}nweger}, \citenamefont {Kohlstedt}, \citenamefont {Kienle}, \citenamefont {Mikolajick}, \citenamefont {Fichtner},\ and\ \citenamefont {Schroeder}}]{Guido_ACSAMI_thermal_2023}%
  \BibitemOpen
  \bibfield  {author} {\bibinfo {author} {\bibfnamefont {R.}~\bibnamefont {Guido}}, \bibinfo {author} {\bibfnamefont {P.~D.}\ \bibnamefont {Lomenzo}}, \bibinfo {author} {\bibfnamefont {M.~R.}\ \bibnamefont {Islam}}, \bibinfo {author} {\bibfnamefont {N.}~\bibnamefont {Wolff}}, \bibinfo {author} {\bibfnamefont {M.}~\bibnamefont {Gremmel}}, \bibinfo {author} {\bibfnamefont {G.}~\bibnamefont {Sch{\"o}nweger}}, \bibinfo {author} {\bibfnamefont {H.}~\bibnamefont {Kohlstedt}}, \bibinfo {author} {\bibfnamefont {L.}~\bibnamefont {Kienle}}, \bibinfo {author} {\bibfnamefont {T.}~\bibnamefont {Mikolajick}}, \bibinfo {author} {\bibfnamefont {S.}~\bibnamefont {Fichtner}},\ and\ \bibinfo {author} {\bibfnamefont {U.}~\bibnamefont {Schroeder}},\ }\href {https://doi.org/10.1021/acsami.2c18313} {\bibfield  {journal} {\bibinfo  {journal} {ACS Applied Materials \& Interfaces}\ }\textbf {\bibinfo {volume} {15}},\ \bibinfo {pages} {7030} (\bibinfo {year} {2023})}\BibitemShut {NoStop}%
\bibitem [{\citenamefont {Wang}\ \emph {et~al.}(2023)\citenamefont {Wang}, \citenamefont {Wang}, \citenamefont {Mondal}, \citenamefont {Hu}, \citenamefont {Liu},\ and\ \citenamefont {Mi}}]{wang2023dawn}%
  \BibitemOpen
  \bibfield  {author} {\bibinfo {author} {\bibfnamefont {P.}~\bibnamefont {Wang}}, \bibinfo {author} {\bibfnamefont {D.}~\bibnamefont {Wang}}, \bibinfo {author} {\bibfnamefont {S.}~\bibnamefont {Mondal}}, \bibinfo {author} {\bibfnamefont {M.}~\bibnamefont {Hu}}, \bibinfo {author} {\bibfnamefont {J.}~\bibnamefont {Liu}},\ and\ \bibinfo {author} {\bibfnamefont {Z.}~\bibnamefont {Mi}},\ }\href {https://doi.org/10.1088/1361-6641/acb80e} {\bibfield  {journal} {\bibinfo  {journal} {Semiconductor Science and Technology}\ }\textbf {\bibinfo {volume} {38}},\ \bibinfo {pages} {043002} (\bibinfo {year} {2023})},\ \bibinfo {note} {publisher: IOP Publishing}\BibitemShut {NoStop}%
\bibitem [{\citenamefont {Kim}\ \emph {et~al.}(2023)\citenamefont {Kim}, \citenamefont {Karpov}, \citenamefont {Olsson},\ and\ \citenamefont {Jariwala}}]{kim2023wurtziteNatNano}%
  \BibitemOpen
  \bibfield  {author} {\bibinfo {author} {\bibfnamefont {K.-H.}\ \bibnamefont {Kim}}, \bibinfo {author} {\bibfnamefont {I.}~\bibnamefont {Karpov}}, \bibinfo {author} {\bibfnamefont {R.~H.}\ \bibnamefont {Olsson}},\ and\ \bibinfo {author} {\bibfnamefont {D.}~\bibnamefont {Jariwala}},\ }\href {https://doi.org/10.1038/s41565-023-01361-y} {\bibfield  {journal} {\bibinfo  {journal} {Nature Nanotechnology}\ }\textbf {\bibinfo {volume} {18}},\ \bibinfo {pages} {422} (\bibinfo {year} {2023})},\ \bibinfo {note} {publisher: Nature Publishing Group}\BibitemShut {NoStop}%
\bibitem [{\citenamefont {Wang}\ \emph {et~al.}(2024)\citenamefont {Wang}, \citenamefont {Yang}, \citenamefont {Liu}, \citenamefont {Wang},\ and\ \citenamefont {Mi}}]{Wang_Mi_APL2024_Perspectives}%
  \BibitemOpen
  \bibfield  {author} {\bibinfo {author} {\bibfnamefont {D.}~\bibnamefont {Wang}}, \bibinfo {author} {\bibfnamefont {S.}~\bibnamefont {Yang}}, \bibinfo {author} {\bibfnamefont {J.}~\bibnamefont {Liu}}, \bibinfo {author} {\bibfnamefont {D.}~\bibnamefont {Wang}},\ and\ \bibinfo {author} {\bibfnamefont {Z.}~\bibnamefont {Mi}},\ }\href {https://doi.org/10.1063/5.0206005} {\bibfield  {journal} {\bibinfo  {journal} {Applied Physics Letters}\ }\textbf {\bibinfo {volume} {124}},\ \bibinfo {pages} {150501} (\bibinfo {year} {2024})}\BibitemShut {NoStop}%
\bibitem [{\citenamefont {Casamento}\ \emph {et~al.}(2024)\citenamefont {Casamento}, \citenamefont {Baksa}, \citenamefont {Behrendt}, \citenamefont {Calderon}, \citenamefont {Goodling}, \citenamefont {Hayden}, \citenamefont {He}, \citenamefont {Jacques}, \citenamefont {Lee}, \citenamefont {Smith}, \citenamefont {Suceava}, \citenamefont {Tran}, \citenamefont {Zheng}, \citenamefont {Zu}, \citenamefont {Beechem}, \citenamefont {Dabo}, \citenamefont {Dickey}, \citenamefont {Esteves}, \citenamefont {Gopalan}, \citenamefont {Henry}, \citenamefont {Ihlefeld}, \citenamefont {Jackson}, \citenamefont {Kalinin}, \citenamefont {Kelley}, \citenamefont {Liu}, \citenamefont {Rappe}, \citenamefont {Redwing}, \citenamefont {Trolier-McKinstry},\ and\ \citenamefont {Maria}}]{Casamento_APL_2024_Perspectives}%
  \BibitemOpen
  \bibfield  {author} {\bibinfo {author} {\bibfnamefont {J.}~\bibnamefont {Casamento}}, \bibinfo {author} {\bibfnamefont {S.~M.}\ \bibnamefont {Baksa}}, \bibinfo {author} {\bibfnamefont {D.}~\bibnamefont {Behrendt}}, \bibinfo {author} {\bibfnamefont {S.}~\bibnamefont {Calderon}}, \bibinfo {author} {\bibfnamefont {D.}~\bibnamefont {Goodling}}, \bibinfo {author} {\bibfnamefont {J.}~\bibnamefont {Hayden}}, \bibinfo {author} {\bibfnamefont {F.}~\bibnamefont {He}}, \bibinfo {author} {\bibfnamefont {L.}~\bibnamefont {Jacques}}, \bibinfo {author} {\bibfnamefont {S.~H.}\ \bibnamefont {Lee}}, \bibinfo {author} {\bibfnamefont {W.}~\bibnamefont {Smith}}, \bibinfo {author} {\bibfnamefont {A.}~\bibnamefont {Suceava}}, \bibinfo {author} {\bibfnamefont {Q.}~\bibnamefont {Tran}}, \bibinfo {author} {\bibfnamefont {X.}~\bibnamefont {Zheng}}, \bibinfo {author} {\bibfnamefont {R.}~\bibnamefont {Zu}}, \bibinfo {author} {\bibfnamefont {T.}~\bibnamefont {Beechem}}, \bibinfo {author} {\bibfnamefont {I.}~\bibnamefont {Dabo}},
  \bibinfo {author} {\bibfnamefont {E.~C.}\ \bibnamefont {Dickey}}, \bibinfo {author} {\bibfnamefont {G.}~\bibnamefont {Esteves}}, \bibinfo {author} {\bibfnamefont {V.}~\bibnamefont {Gopalan}}, \bibinfo {author} {\bibfnamefont {M.~D.}\ \bibnamefont {Henry}}, \bibinfo {author} {\bibfnamefont {J.~F.}\ \bibnamefont {Ihlefeld}}, \bibinfo {author} {\bibfnamefont {T.~N.}\ \bibnamefont {Jackson}}, \bibinfo {author} {\bibfnamefont {S.~V.}\ \bibnamefont {Kalinin}}, \bibinfo {author} {\bibfnamefont {K.~P.}\ \bibnamefont {Kelley}}, \bibinfo {author} {\bibfnamefont {Y.}~\bibnamefont {Liu}}, \bibinfo {author} {\bibfnamefont {A.~M.}\ \bibnamefont {Rappe}}, \bibinfo {author} {\bibfnamefont {J.}~\bibnamefont {Redwing}}, \bibinfo {author} {\bibfnamefont {S.}~\bibnamefont {Trolier-McKinstry}},\ and\ \bibinfo {author} {\bibfnamefont {J.-P.}\ \bibnamefont {Maria}},\ }\href {https://doi.org/10.1063/5.0185066} {\bibfield  {journal} {\bibinfo  {journal} {Applied Physics Letters}\ }\textbf {\bibinfo {volume} {124}},\ \bibinfo
  {pages} {080501} (\bibinfo {year} {2024})}\BibitemShut {NoStop}%
\bibitem [{\citenamefont {Wang}\ \emph {et~al.}(2025{\natexlab{a}})\citenamefont {Wang}, \citenamefont {Wang}, \citenamefont {Molla}, \citenamefont {Liu}, \citenamefont {Yang}, \citenamefont {Yuan}, \citenamefont {Liu}, \citenamefont {Hu}, \citenamefont {Wu}, \citenamefont {Ma}, \citenamefont {Sun}, \citenamefont {Guo}, \citenamefont {Kioupakis},\ and\ \citenamefont {Mi}}]{wang2025electric}%
  \BibitemOpen
  \bibfield  {author} {\bibinfo {author} {\bibfnamefont {D.}~\bibnamefont {Wang}}, \bibinfo {author} {\bibfnamefont {D.}~\bibnamefont {Wang}}, \bibinfo {author} {\bibfnamefont {M.}~\bibnamefont {Molla}}, \bibinfo {author} {\bibfnamefont {Y.}~\bibnamefont {Liu}}, \bibinfo {author} {\bibfnamefont {S.}~\bibnamefont {Yang}}, \bibinfo {author} {\bibfnamefont {S.}~\bibnamefont {Yuan}}, \bibinfo {author} {\bibfnamefont {J.}~\bibnamefont {Liu}}, \bibinfo {author} {\bibfnamefont {M.}~\bibnamefont {Hu}}, \bibinfo {author} {\bibfnamefont {Y.}~\bibnamefont {Wu}}, \bibinfo {author} {\bibfnamefont {T.}~\bibnamefont {Ma}}, \bibinfo {author} {\bibfnamefont {K.}~\bibnamefont {Sun}}, \bibinfo {author} {\bibfnamefont {H.}~\bibnamefont {Guo}}, \bibinfo {author} {\bibfnamefont {E.}~\bibnamefont {Kioupakis}},\ and\ \bibinfo {author} {\bibfnamefont {Z.}~\bibnamefont {Mi}},\ }\href {https://doi.org/10.1038/s41586-025-08812-7} {\bibfield  {journal} {\bibinfo  {journal} {Nature}\ }\textbf {\bibinfo {volume} {641}},\ \bibinfo {pages}
  {76} (\bibinfo {year} {2025}{\natexlab{a}})}\BibitemShut {NoStop}%
\bibitem [{\citenamefont {Li}\ \emph {et~al.}(2025)\citenamefont {Li}, \citenamefont {Wang}, \citenamefont {Gao}, \citenamefont {Damjanovic}, \citenamefont {Chen},\ and\ \citenamefont {Zhang}}]{Li_science_2025_review}%
  \BibitemOpen
  \bibfield  {author} {\bibinfo {author} {\bibfnamefont {F.}~\bibnamefont {Li}}, \bibinfo {author} {\bibfnamefont {B.}~\bibnamefont {Wang}}, \bibinfo {author} {\bibfnamefont {X.}~\bibnamefont {Gao}}, \bibinfo {author} {\bibfnamefont {D.}~\bibnamefont {Damjanovic}}, \bibinfo {author} {\bibfnamefont {L.-Q.}\ \bibnamefont {Chen}},\ and\ \bibinfo {author} {\bibfnamefont {S.}~\bibnamefont {Zhang}},\ }\href {https://doi.org/10.1126/science.adn4926} {\bibfield  {journal} {\bibinfo  {journal} {Science}\ }\textbf {\bibinfo {volume} {389}},\ \bibinfo {pages} {eadn4926} (\bibinfo {year} {2025})}\BibitemShut {NoStop}%
\bibitem [{\citenamefont {Fichtner}\ \emph {et~al.}(2025)\citenamefont {Fichtner}, \citenamefont {Schönweger}, \citenamefont {Lee}, \citenamefont {Yazawa}, \citenamefont {Gorai},\ and\ \citenamefont {Brennecka}}]{Fichtner_2025_APR}%
  \BibitemOpen
  \bibfield  {author} {\bibinfo {author} {\bibfnamefont {S.}~\bibnamefont {Fichtner}}, \bibinfo {author} {\bibfnamefont {G.}~\bibnamefont {Schönweger}}, \bibinfo {author} {\bibfnamefont {C.-W.}\ \bibnamefont {Lee}}, \bibinfo {author} {\bibfnamefont {K.}~\bibnamefont {Yazawa}}, \bibinfo {author} {\bibfnamefont {P.}~\bibnamefont {Gorai}},\ and\ \bibinfo {author} {\bibfnamefont {G.~L.}\ \bibnamefont {Brennecka}},\ }\href {https://doi.org/10.1063/5.0249265} {\bibfield  {journal} {\bibinfo  {journal} {Applied Physics Reviews}\ }\textbf {\bibinfo {volume} {12}},\ \bibinfo {pages} {021310} (\bibinfo {year} {2025})}\BibitemShut {NoStop}%
\bibitem [{\citenamefont {Cho}\ \emph {et~al.}(2026)\citenamefont {Cho}, \citenamefont {Wang}, \citenamefont {Leblanc}, \citenamefont {Zhang}, \citenamefont {He}, \citenamefont {Han}, \citenamefont {Tong}, \citenamefont {Bulumulla}, \citenamefont {Tan}, \citenamefont {Olsson},\ and\ \citenamefont {Jariwala}}]{cho_write_NC_2026}%
  \BibitemOpen
  \bibfield  {author} {\bibinfo {author} {\bibfnamefont {H.}~\bibnamefont {Cho}}, \bibinfo {author} {\bibfnamefont {Y.}~\bibnamefont {Wang}}, \bibinfo {author} {\bibfnamefont {C.}~\bibnamefont {Leblanc}}, \bibinfo {author} {\bibfnamefont {Y.}~\bibnamefont {Zhang}}, \bibinfo {author} {\bibfnamefont {Y.}~\bibnamefont {He}}, \bibinfo {author} {\bibfnamefont {Z.}~\bibnamefont {Han}}, \bibinfo {author} {\bibfnamefont {X.}~\bibnamefont {Tong}}, \bibinfo {author} {\bibfnamefont {V.~D.}\ \bibnamefont {Bulumulla}}, \bibinfo {author} {\bibfnamefont {J.}~\bibnamefont {Tan}}, \bibinfo {author} {\bibfnamefont {R.~H.}\ \bibnamefont {Olsson}},\ and\ \bibinfo {author} {\bibfnamefont {D.}~\bibnamefont {Jariwala}},\ }\href {https://doi.org/10.1038/s41467-025-68221-2} {\bibfield  {journal} {\bibinfo  {journal} {Nature Communications}\ }\textbf {\bibinfo {volume} {17}},\ \bibinfo {pages} {1507} (\bibinfo {year} {2026})}\BibitemShut {NoStop}%
\bibitem [{\citenamefont {Messi}\ \emph {et~al.}(2025)\citenamefont {Messi}, \citenamefont {Patidar}, \citenamefont {Rodkey}, \citenamefont {Dräyer}, \citenamefont {Trassin},\ and\ \citenamefont {Siol}}]{Messi_APLM_2025}%
  \BibitemOpen
  \bibfield  {author} {\bibinfo {author} {\bibfnamefont {F.}~\bibnamefont {Messi}}, \bibinfo {author} {\bibfnamefont {J.}~\bibnamefont {Patidar}}, \bibinfo {author} {\bibfnamefont {N.}~\bibnamefont {Rodkey}}, \bibinfo {author} {\bibfnamefont {C.~W.}\ \bibnamefont {Dräyer}}, \bibinfo {author} {\bibfnamefont {M.}~\bibnamefont {Trassin}},\ and\ \bibinfo {author} {\bibfnamefont {S.}~\bibnamefont {Siol}},\ }\href {https://doi.org/10.1063/5.0267904} {\bibfield  {journal} {\bibinfo  {journal} {APL Materials}\ }\textbf {\bibinfo {volume} {13}},\ \bibinfo {pages} {051123} (\bibinfo {year} {2025})}\BibitemShut {NoStop}%
\bibitem [{\citenamefont {Hirata}\ \emph {et~al.}(2025)\citenamefont {Hirata}, \citenamefont {Niitsu}, \citenamefont {Anggraini}, \citenamefont {Kageura}, \citenamefont {Uehara}, \citenamefont {Yamada},\ and\ \citenamefont {Akiyama}}]{HIRATA_MT_2025}%
  \BibitemOpen
  \bibfield  {author} {\bibinfo {author} {\bibfnamefont {K.}~\bibnamefont {Hirata}}, \bibinfo {author} {\bibfnamefont {K.}~\bibnamefont {Niitsu}}, \bibinfo {author} {\bibfnamefont {S.~A.}\ \bibnamefont {Anggraini}}, \bibinfo {author} {\bibfnamefont {T.}~\bibnamefont {Kageura}}, \bibinfo {author} {\bibfnamefont {M.}~\bibnamefont {Uehara}}, \bibinfo {author} {\bibfnamefont {H.}~\bibnamefont {Yamada}},\ and\ \bibinfo {author} {\bibfnamefont {M.}~\bibnamefont {Akiyama}},\ }\href {https://doi.org/https://doi.org/10.1016/j.mattod.2024.12.011} {\bibfield  {journal} {\bibinfo  {journal} {Materials Today}\ }\textbf {\bibinfo {volume} {83}},\ \bibinfo {pages} {85} (\bibinfo {year} {2025})}\BibitemShut {NoStop}%
\bibitem [{\citenamefont {Behrendt}\ \emph {et~al.}(2026)\citenamefont {Behrendt}, \citenamefont {Samanta},\ and\ \citenamefont {Rappe}}]{Behrendt_PRL_2026_Fractals}%
  \BibitemOpen
  \bibfield  {author} {\bibinfo {author} {\bibfnamefont {D.}~\bibnamefont {Behrendt}}, \bibinfo {author} {\bibfnamefont {A.}~\bibnamefont {Samanta}},\ and\ \bibinfo {author} {\bibfnamefont {A.~M.}\ \bibnamefont {Rappe}},\ }\href {https://doi.org/10.1103/2qs8-yxmr} {\bibfield  {journal} {\bibinfo  {journal} {Phys. Rev. Lett.}\ }\textbf {\bibinfo {volume} {136}},\ \bibinfo {pages} {176101} (\bibinfo {year} {2026})}\BibitemShut {NoStop}%
\bibitem [{\citenamefont {Tasn\'adi}\ \emph {et~al.}(2010)\citenamefont {Tasn\'adi}, \citenamefont {Alling}, \citenamefont {H\"oglund}, \citenamefont {Wingqvist}, \citenamefont {Birch}, \citenamefont {Hultman},\ and\ \citenamefont {Abrikosov}}]{Tasnadi-PRL-2010-origin}%
  \BibitemOpen
  \bibfield  {author} {\bibinfo {author} {\bibfnamefont {F.}~\bibnamefont {Tasn\'adi}}, \bibinfo {author} {\bibfnamefont {B.}~\bibnamefont {Alling}}, \bibinfo {author} {\bibfnamefont {C.}~\bibnamefont {H\"oglund}}, \bibinfo {author} {\bibfnamefont {G.}~\bibnamefont {Wingqvist}}, \bibinfo {author} {\bibfnamefont {J.}~\bibnamefont {Birch}}, \bibinfo {author} {\bibfnamefont {L.}~\bibnamefont {Hultman}},\ and\ \bibinfo {author} {\bibfnamefont {I.~A.}\ \bibnamefont {Abrikosov}},\ }\href {https://doi.org/10.1103/PhysRevLett.104.137601} {\bibfield  {journal} {\bibinfo  {journal} {Phys. Rev. Lett.}\ }\textbf {\bibinfo {volume} {104}},\ \bibinfo {pages} {137601} (\bibinfo {year} {2010})}\BibitemShut {NoStop}%
\bibitem [{\citenamefont {Zhang}\ \emph {et~al.}(2013)\citenamefont {Zhang}, \citenamefont {Holec}, \citenamefont {Fu}, \citenamefont {Humphreys},\ and\ \citenamefont {Moram}}]{zhang2013tunable}%
  \BibitemOpen
  \bibfield  {author} {\bibinfo {author} {\bibfnamefont {S.}~\bibnamefont {Zhang}}, \bibinfo {author} {\bibfnamefont {D.}~\bibnamefont {Holec}}, \bibinfo {author} {\bibfnamefont {W.~Y.}\ \bibnamefont {Fu}}, \bibinfo {author} {\bibfnamefont {C.~J.}\ \bibnamefont {Humphreys}},\ and\ \bibinfo {author} {\bibfnamefont {M.~A.}\ \bibnamefont {Moram}},\ }\href {https://doi.org/10.1063/1.4824179} {\bibfield  {journal} {\bibinfo  {journal} {Journal of Applied Physics}\ }\textbf {\bibinfo {volume} {114}},\ \bibinfo {pages} {133510} (\bibinfo {year} {2013})}\BibitemShut {NoStop}%
\bibitem [{\citenamefont {Caro}\ \emph {et~al.}(2015)\citenamefont {Caro}, \citenamefont {Zhang}, \citenamefont {Riekkinen}, \citenamefont {Ylilammi}, \citenamefont {Moram}, \citenamefont {Lopez-Acevedo}, \citenamefont {Molarius},\ and\ \citenamefont {Laurila}}]{Caro_JPCM-2015-Piezoelectric}%
  \BibitemOpen
  \bibfield  {author} {\bibinfo {author} {\bibfnamefont {M.~A.}\ \bibnamefont {Caro}}, \bibinfo {author} {\bibfnamefont {S.}~\bibnamefont {Zhang}}, \bibinfo {author} {\bibfnamefont {T.}~\bibnamefont {Riekkinen}}, \bibinfo {author} {\bibfnamefont {M.}~\bibnamefont {Ylilammi}}, \bibinfo {author} {\bibfnamefont {M.~A.}\ \bibnamefont {Moram}}, \bibinfo {author} {\bibfnamefont {O.}~\bibnamefont {Lopez-Acevedo}}, \bibinfo {author} {\bibfnamefont {J.}~\bibnamefont {Molarius}},\ and\ \bibinfo {author} {\bibfnamefont {T.}~\bibnamefont {Laurila}},\ }\href {https://doi.org/10.1088/0953-8984/27/24/245901} {\bibfield  {journal} {\bibinfo  {journal} {Journal of Physics: Condensed Matter}\ }\textbf {\bibinfo {volume} {27}},\ \bibinfo {pages} {245901} (\bibinfo {year} {2015})}\BibitemShut {NoStop}%
\bibitem [{\citenamefont {Talley}\ \emph {et~al.}(2018)\citenamefont {Talley}, \citenamefont {Millican}, \citenamefont {Mangum}, \citenamefont {Siol}, \citenamefont {Musgrave}, \citenamefont {Gorman}, \citenamefont {Holder}, \citenamefont {Zakutayev},\ and\ \citenamefont {Brennecka}}]{Talley-2018-PRM-implications}%
  \BibitemOpen
  \bibfield  {author} {\bibinfo {author} {\bibfnamefont {K.~R.}\ \bibnamefont {Talley}}, \bibinfo {author} {\bibfnamefont {S.~L.}\ \bibnamefont {Millican}}, \bibinfo {author} {\bibfnamefont {J.}~\bibnamefont {Mangum}}, \bibinfo {author} {\bibfnamefont {S.}~\bibnamefont {Siol}}, \bibinfo {author} {\bibfnamefont {C.~B.}\ \bibnamefont {Musgrave}}, \bibinfo {author} {\bibfnamefont {B.}~\bibnamefont {Gorman}}, \bibinfo {author} {\bibfnamefont {A.~M.}\ \bibnamefont {Holder}}, \bibinfo {author} {\bibfnamefont {A.}~\bibnamefont {Zakutayev}},\ and\ \bibinfo {author} {\bibfnamefont {G.~L.}\ \bibnamefont {Brennecka}},\ }\href {https://doi.org/10.1103/PhysRevMaterials.2.063802} {\bibfield  {journal} {\bibinfo  {journal} {Phys. Rev. Mater.}\ }\textbf {\bibinfo {volume} {2}},\ \bibinfo {pages} {063802} (\bibinfo {year} {2018})}\BibitemShut {NoStop}%
\bibitem [{\citenamefont {Kyrtsos}\ \emph {et~al.}(2019)\citenamefont {Kyrtsos}, \citenamefont {Matsubara},\ and\ \citenamefont {Bellotti}}]{Kyrtsos_2019_PRB_first}%
  \BibitemOpen
  \bibfield  {author} {\bibinfo {author} {\bibfnamefont {A.}~\bibnamefont {Kyrtsos}}, \bibinfo {author} {\bibfnamefont {M.}~\bibnamefont {Matsubara}},\ and\ \bibinfo {author} {\bibfnamefont {E.}~\bibnamefont {Bellotti}},\ }\href {https://doi.org/10.1103/PhysRevB.99.035201} {\bibfield  {journal} {\bibinfo  {journal} {Phys. Rev. B}\ }\textbf {\bibinfo {volume} {99}},\ \bibinfo {pages} {035201} (\bibinfo {year} {2019})}\BibitemShut {NoStop}%
\bibitem [{\citenamefont {Moriwake}\ \emph {et~al.}(2020)\citenamefont {Moriwake}, \citenamefont {Yokoi}, \citenamefont {Taguchi}, \citenamefont {Ogawa}, \citenamefont {Fisher}, \citenamefont {Kuwabara}, \citenamefont {Sato}, \citenamefont {Shimizu}, \citenamefont {Hamasaki}, \citenamefont {Takashima},\ and\ \citenamefont {Itoh}}]{Moriwake_2020_APLM_computational}%
  \BibitemOpen
  \bibfield  {author} {\bibinfo {author} {\bibfnamefont {H.}~\bibnamefont {Moriwake}}, \bibinfo {author} {\bibfnamefont {R.}~\bibnamefont {Yokoi}}, \bibinfo {author} {\bibfnamefont {A.}~\bibnamefont {Taguchi}}, \bibinfo {author} {\bibfnamefont {T.}~\bibnamefont {Ogawa}}, \bibinfo {author} {\bibfnamefont {C.~A.~J.}\ \bibnamefont {Fisher}}, \bibinfo {author} {\bibfnamefont {A.}~\bibnamefont {Kuwabara}}, \bibinfo {author} {\bibfnamefont {Y.}~\bibnamefont {Sato}}, \bibinfo {author} {\bibfnamefont {T.}~\bibnamefont {Shimizu}}, \bibinfo {author} {\bibfnamefont {Y.}~\bibnamefont {Hamasaki}}, \bibinfo {author} {\bibfnamefont {H.}~\bibnamefont {Takashima}},\ and\ \bibinfo {author} {\bibfnamefont {M.}~\bibnamefont {Itoh}},\ }\href {https://doi.org/10.1063/5.0023626} {\bibfield  {journal} {\bibinfo  {journal} {APL Materials}\ }\textbf {\bibinfo {volume} {8}},\ \bibinfo {pages} {121102} (\bibinfo {year} {2020})}\BibitemShut {NoStop}%
\bibitem [{\citenamefont {Wang}\ \emph {et~al.}(2021{\natexlab{b}})\citenamefont {Wang}, \citenamefont {Adamski}, \citenamefont {Mu},\ and\ \citenamefont {Van~de Walle}}]{Wang_2021_JAP_Piezoelectric}%
  \BibitemOpen
  \bibfield  {author} {\bibinfo {author} {\bibfnamefont {H.}~\bibnamefont {Wang}}, \bibinfo {author} {\bibfnamefont {N.}~\bibnamefont {Adamski}}, \bibinfo {author} {\bibfnamefont {S.}~\bibnamefont {Mu}},\ and\ \bibinfo {author} {\bibfnamefont {C.~G.}\ \bibnamefont {Van~de Walle}},\ }\href {https://doi.org/10.1063/5.0056485} {\bibfield  {journal} {\bibinfo  {journal} {Journal of Applied Physics}\ }\textbf {\bibinfo {volume} {130}},\ \bibinfo {pages} {104101} (\bibinfo {year} {2021}{\natexlab{b}})}\BibitemShut {NoStop}%
\bibitem [{\citenamefont {Liu}\ \emph {et~al.}(2021)\citenamefont {Liu}, \citenamefont {Wang}, \citenamefont {Yang}, \citenamefont {Cao}, \citenamefont {Chen}, \citenamefont {Li}, \citenamefont {Liu}, \citenamefont {Loke}, \citenamefont {Kang},\ and\ \citenamefont {Zhu}}]{liu_multiscale_2021}%
  \BibitemOpen
  \bibfield  {author} {\bibinfo {author} {\bibfnamefont {C.}~\bibnamefont {Liu}}, \bibinfo {author} {\bibfnamefont {Q.}~\bibnamefont {Wang}}, \bibinfo {author} {\bibfnamefont {W.}~\bibnamefont {Yang}}, \bibinfo {author} {\bibfnamefont {T.}~\bibnamefont {Cao}}, \bibinfo {author} {\bibfnamefont {L.}~\bibnamefont {Chen}}, \bibinfo {author} {\bibfnamefont {M.}~\bibnamefont {Li}}, \bibinfo {author} {\bibfnamefont {F.}~\bibnamefont {Liu}}, \bibinfo {author} {\bibfnamefont {D.~K.}\ \bibnamefont {Loke}}, \bibinfo {author} {\bibfnamefont {J.}~\bibnamefont {Kang}},\ and\ \bibinfo {author} {\bibfnamefont {Y.}~\bibnamefont {Zhu}},\ }in\ \href {https://doi.org/10.1109/IEDM19574.2021.9720535} {\emph {\bibinfo {booktitle} {2021 {IEEE} {International} {Electron} {Devices} {Meeting} ({IEDM})}}}\ (\bibinfo {year} {2021})\ pp.\ \bibinfo {pages} {8.1.1--8.1.4},\ \bibinfo {note} {iSSN: 2156-017X}\BibitemShut {NoStop}%
\bibitem [{\citenamefont {Furuta}\ \emph {et~al.}(2021)\citenamefont {Furuta}, \citenamefont {Hirata}, \citenamefont {Anggraini}, \citenamefont {Akiyama}, \citenamefont {Uehara},\ and\ \citenamefont {Yamada}}]{Furuta_2021_JAP_first}%
  \BibitemOpen
  \bibfield  {author} {\bibinfo {author} {\bibfnamefont {K.}~\bibnamefont {Furuta}}, \bibinfo {author} {\bibfnamefont {K.}~\bibnamefont {Hirata}}, \bibinfo {author} {\bibfnamefont {S.~A.}\ \bibnamefont {Anggraini}}, \bibinfo {author} {\bibfnamefont {M.}~\bibnamefont {Akiyama}}, \bibinfo {author} {\bibfnamefont {M.}~\bibnamefont {Uehara}},\ and\ \bibinfo {author} {\bibfnamefont {H.}~\bibnamefont {Yamada}},\ }\href {https://doi.org/10.1063/5.0051557} {\bibfield  {journal} {\bibinfo  {journal} {Journal of Applied Physics}\ }\textbf {\bibinfo {volume} {130}},\ \bibinfo {pages} {024104} (\bibinfo {year} {2021})}\BibitemShut {NoStop}%
\bibitem [{\citenamefont {Urban}\ \emph {et~al.}(2021)\citenamefont {Urban}, \citenamefont {Ambacher},\ and\ \citenamefont {Els\"asser}}]{Urban_2021_PRB_first}%
  \BibitemOpen
  \bibfield  {author} {\bibinfo {author} {\bibfnamefont {D.~F.}\ \bibnamefont {Urban}}, \bibinfo {author} {\bibfnamefont {O.}~\bibnamefont {Ambacher}},\ and\ \bibinfo {author} {\bibfnamefont {C.}~\bibnamefont {Els\"asser}},\ }\href {https://doi.org/10.1103/PhysRevB.103.115204} {\bibfield  {journal} {\bibinfo  {journal} {Phys. Rev. B}\ }\textbf {\bibinfo {volume} {103}},\ \bibinfo {pages} {115204} (\bibinfo {year} {2021})}\BibitemShut {NoStop}%
\bibitem [{\citenamefont {Ye}\ \emph {et~al.}(2021)\citenamefont {Ye}, \citenamefont {Han}, \citenamefont {Yeu}, \citenamefont {Hwang},\ and\ \citenamefont {Choi}}]{Ye_2021_PSSR_atomistic}%
  \BibitemOpen
  \bibfield  {author} {\bibinfo {author} {\bibfnamefont {K.~H.}\ \bibnamefont {Ye}}, \bibinfo {author} {\bibfnamefont {G.}~\bibnamefont {Han}}, \bibinfo {author} {\bibfnamefont {I.~W.}\ \bibnamefont {Yeu}}, \bibinfo {author} {\bibfnamefont {C.~S.}\ \bibnamefont {Hwang}},\ and\ \bibinfo {author} {\bibfnamefont {J.-H.}\ \bibnamefont {Choi}},\ }\href {https://doi.org/https://doi.org/10.1002/pssr.202100009} {\bibfield  {journal} {\bibinfo  {journal} {physica status solidi (RRL) – Rapid Research Letters}\ }\textbf {\bibinfo {volume} {15}},\ \bibinfo {pages} {2100009} (\bibinfo {year} {2021})}\BibitemShut {NoStop}%
\bibitem [{\citenamefont {Wang}\ \emph {et~al.}(2022)\citenamefont {Wang}, \citenamefont {Go}, \citenamefont {Liu}, \citenamefont {Li}, \citenamefont {Zhu}, \citenamefont {Li}, \citenamefont {Lee},\ and\ \citenamefont {Loke}}]{Wang_2022_AIPA_understanding}%
  \BibitemOpen
  \bibfield  {author} {\bibinfo {author} {\bibfnamefont {Q.}~\bibnamefont {Wang}}, \bibinfo {author} {\bibfnamefont {S.-X.}\ \bibnamefont {Go}}, \bibinfo {author} {\bibfnamefont {C.}~\bibnamefont {Liu}}, \bibinfo {author} {\bibfnamefont {M.}~\bibnamefont {Li}}, \bibinfo {author} {\bibfnamefont {Y.}~\bibnamefont {Zhu}}, \bibinfo {author} {\bibfnamefont {L.}~\bibnamefont {Li}}, \bibinfo {author} {\bibfnamefont {T.~H.}\ \bibnamefont {Lee}},\ and\ \bibinfo {author} {\bibfnamefont {D.~K.}\ \bibnamefont {Loke}},\ }\href {https://doi.org/10.1063/5.0126651} {\bibfield  {journal} {\bibinfo  {journal} {AIP Advances}\ }\textbf {\bibinfo {volume} {12}},\ \bibinfo {pages} {125303} (\bibinfo {year} {2022})}\BibitemShut {NoStop}%
\bibitem [{\citenamefont {Balestra}\ \emph {et~al.}(2022)\citenamefont {Balestra}, \citenamefont {Gnani},\ and\ \citenamefont {Reggiani}}]{Balestra_2022_JAP_electron}%
  \BibitemOpen
  \bibfield  {author} {\bibinfo {author} {\bibfnamefont {L.}~\bibnamefont {Balestra}}, \bibinfo {author} {\bibfnamefont {E.}~\bibnamefont {Gnani}},\ and\ \bibinfo {author} {\bibfnamefont {S.}~\bibnamefont {Reggiani}},\ }\href {https://doi.org/10.1063/5.0115512} {\bibfield  {journal} {\bibinfo  {journal} {Journal of Applied Physics}\ }\textbf {\bibinfo {volume} {132}},\ \bibinfo {pages} {215108} (\bibinfo {year} {2022})}\BibitemShut {NoStop}%
\bibitem [{\citenamefont {Yazawa}\ \emph {et~al.}(2022)\citenamefont {Yazawa}, \citenamefont {S.~Mangum}, \citenamefont {Gorai}, \citenamefont {L.~Brennecka},\ and\ \citenamefont {Zakutayev}}]{yazawa_JMCC_2022_local}%
  \BibitemOpen
  \bibfield  {author} {\bibinfo {author} {\bibfnamefont {K.}~\bibnamefont {Yazawa}}, \bibinfo {author} {\bibfnamefont {J.}~\bibnamefont {S.~Mangum}}, \bibinfo {author} {\bibfnamefont {P.}~\bibnamefont {Gorai}}, \bibinfo {author} {\bibfnamefont {G.}~\bibnamefont {L.~Brennecka}},\ and\ \bibinfo {author} {\bibfnamefont {A.}~\bibnamefont {Zakutayev}},\ }\href {https://doi.org/10.1039/D2TC02682A} {\bibfield  {journal} {\bibinfo  {journal} {Journal of Materials Chemistry C}\ }\textbf {\bibinfo {volume} {10}},\ \bibinfo {pages} {17557} (\bibinfo {year} {2022})}\BibitemShut {NoStop}%
\bibitem [{\citenamefont {Wang}\ \emph {et~al.}(2025{\natexlab{b}})\citenamefont {Wang}, \citenamefont {Mu},\ and\ \citenamefont {Van~de Walle}}]{Wang_APL_2024_towards}%
  \BibitemOpen
  \bibfield  {author} {\bibinfo {author} {\bibfnamefont {H.}~\bibnamefont {Wang}}, \bibinfo {author} {\bibfnamefont {S.}~\bibnamefont {Mu}},\ and\ \bibinfo {author} {\bibfnamefont {C.~G.}\ \bibnamefont {Van~de Walle}},\ }\href {https://doi.org/10.1063/5.0244434} {\bibfield  {journal} {\bibinfo  {journal} {Applied Physics Letters}\ }\textbf {\bibinfo {volume} {126}},\ \bibinfo {pages} {041901} (\bibinfo {year} {2025}{\natexlab{b}})}\BibitemShut {NoStop}%
\bibitem [{\citenamefont {Lee}\ \emph {et~al.}(2024)\citenamefont {Lee}, \citenamefont {Yazawa}, \citenamefont {Zakutayev}, \citenamefont {Brennecka},\ and\ \citenamefont {Gorai}}]{Lee_sciadv_2024_switching}%
  \BibitemOpen
  \bibfield  {author} {\bibinfo {author} {\bibfnamefont {C.-W.}\ \bibnamefont {Lee}}, \bibinfo {author} {\bibfnamefont {K.}~\bibnamefont {Yazawa}}, \bibinfo {author} {\bibfnamefont {A.}~\bibnamefont {Zakutayev}}, \bibinfo {author} {\bibfnamefont {G.~L.}\ \bibnamefont {Brennecka}},\ and\ \bibinfo {author} {\bibfnamefont {P.}~\bibnamefont {Gorai}},\ }\href {https://doi.org/10.1126/sciadv.adl0848} {\bibfield  {journal} {\bibinfo  {journal} {Science Advances}\ }\textbf {\bibinfo {volume} {10}},\ \bibinfo {pages} {eadl0848} (\bibinfo {year} {2024})}\BibitemShut {NoStop}%
\bibitem [{\citenamefont {Pike}\ \emph {et~al.}(2025)\citenamefont {Pike}, \citenamefont {Pachter}, \citenamefont {Kennedy},\ and\ \citenamefont {Glavin}}]{Pike_PRB_2025_understanding}%
  \BibitemOpen
  \bibfield  {author} {\bibinfo {author} {\bibfnamefont {N.~A.}\ \bibnamefont {Pike}}, \bibinfo {author} {\bibfnamefont {R.}~\bibnamefont {Pachter}}, \bibinfo {author} {\bibfnamefont {W.~J.}\ \bibnamefont {Kennedy}},\ and\ \bibinfo {author} {\bibfnamefont {N.}~\bibnamefont {Glavin}},\ }\href {https://doi.org/10.1103/ngv8-vq2k} {\bibfield  {journal} {\bibinfo  {journal} {Phys. Rev. B}\ }\textbf {\bibinfo {volume} {112}},\ \bibinfo {pages} {085204} (\bibinfo {year} {2025})}\BibitemShut {NoStop}%
\bibitem [{\citenamefont {Wang}\ \emph {et~al.}(2025{\natexlab{c}})\citenamefont {Wang}, \citenamefont {Matsubara},\ and\ \citenamefont {Bellotti}}]{Wang_PRM_2025_structural}%
  \BibitemOpen
  \bibfield  {author} {\bibinfo {author} {\bibfnamefont {C.~J.}\ \bibnamefont {Wang}}, \bibinfo {author} {\bibfnamefont {M.}~\bibnamefont {Matsubara}},\ and\ \bibinfo {author} {\bibfnamefont {E.}~\bibnamefont {Bellotti}},\ }\href {https://doi.org/10.1103/z4bq-7xr7} {\bibfield  {journal} {\bibinfo  {journal} {Phys. Rev. Mater.}\ }\textbf {\bibinfo {volume} {9}},\ \bibinfo {pages} {124603} (\bibinfo {year} {2025}{\natexlab{c}})}\BibitemShut {NoStop}%
\bibitem [{\citenamefont {Chen}\ \emph {et~al.}(2025)\citenamefont {Chen}, \citenamefont {Wang}, \citenamefont {Tejerina}, \citenamefont {Yazawa}, \citenamefont {Zakutayev}, \citenamefont {Paillard},\ and\ \citenamefont {Bellaiche}}]{Chen_PRM_2025_towards}%
  \BibitemOpen
  \bibfield  {author} {\bibinfo {author} {\bibfnamefont {P.}~\bibnamefont {Chen}}, \bibinfo {author} {\bibfnamefont {D.}~\bibnamefont {Wang}}, \bibinfo {author} {\bibfnamefont {A.~M.}\ \bibnamefont {Tejerina}}, \bibinfo {author} {\bibfnamefont {K.}~\bibnamefont {Yazawa}}, \bibinfo {author} {\bibfnamefont {A.}~\bibnamefont {Zakutayev}}, \bibinfo {author} {\bibfnamefont {C.}~\bibnamefont {Paillard}},\ and\ \bibinfo {author} {\bibfnamefont {L.}~\bibnamefont {Bellaiche}},\ }\href {https://doi.org/10.1103/9nv5-ryqr} {\bibfield  {journal} {\bibinfo  {journal} {Phys. Rev. Mater.}\ }\textbf {\bibinfo {volume} {9}},\ \bibinfo {pages} {124418} (\bibinfo {year} {2025})}\BibitemShut {NoStop}%
\bibitem [{\citenamefont {Safaltin}\ \emph {et~al.}(2026)\citenamefont {Safaltin}, \citenamefont {Nayak},\ and\ \citenamefont {Alpay}}]{Safaltin_Acta_2026_tailoring}%
  \BibitemOpen
  \bibfield  {author} {\bibinfo {author} {\bibfnamefont {S.}~\bibnamefont {Safaltin}}, \bibinfo {author} {\bibfnamefont {S.~K.}\ \bibnamefont {Nayak}},\ and\ \bibinfo {author} {\bibfnamefont {S.}~\bibnamefont {Alpay}},\ }\href {https://doi.org/https://doi.org/10.1016/j.actamat.2026.122040} {\bibfield  {journal} {\bibinfo  {journal} {Acta Materialia}\ }\textbf {\bibinfo {volume} {308}},\ \bibinfo {pages} {122040} (\bibinfo {year} {2026})}\BibitemShut {NoStop}%
\bibitem [{\citenamefont {Zunger}\ \emph {et~al.}(1990)\citenamefont {Zunger}, \citenamefont {Wei}, \citenamefont {Ferreira},\ and\ \citenamefont {Bernard}}]{Zunger_PRL_1990_SQS}%
  \BibitemOpen
  \bibfield  {author} {\bibinfo {author} {\bibfnamefont {A.}~\bibnamefont {Zunger}}, \bibinfo {author} {\bibfnamefont {S.-H.}\ \bibnamefont {Wei}}, \bibinfo {author} {\bibfnamefont {L.~G.}\ \bibnamefont {Ferreira}},\ and\ \bibinfo {author} {\bibfnamefont {J.~E.}\ \bibnamefont {Bernard}},\ }\href {https://doi.org/10.1103/PhysRevLett.65.353} {\bibfield  {journal} {\bibinfo  {journal} {Phys. Rev. Lett.}\ }\textbf {\bibinfo {volume} {65}},\ \bibinfo {pages} {353} (\bibinfo {year} {1990})}\BibitemShut {NoStop}%
\bibitem [{\citenamefont {Ji}\ \emph {et~al.}(2019)\citenamefont {Ji}, \citenamefont {Urban}, \citenamefont {Kitchaev}, \citenamefont {Kwon}, \citenamefont {Artrith}, \citenamefont {Ophus}, \citenamefont {Huang}, \citenamefont {Cai}, \citenamefont {Shi}, \citenamefont {Kim}, \citenamefont {Kim},\ and\ \citenamefont {Ceder}}]{ji_hidden_NC_2019}%
  \BibitemOpen
  \bibfield  {author} {\bibinfo {author} {\bibfnamefont {H.}~\bibnamefont {Ji}}, \bibinfo {author} {\bibfnamefont {A.}~\bibnamefont {Urban}}, \bibinfo {author} {\bibfnamefont {D.~A.}\ \bibnamefont {Kitchaev}}, \bibinfo {author} {\bibfnamefont {D.-H.}\ \bibnamefont {Kwon}}, \bibinfo {author} {\bibfnamefont {N.}~\bibnamefont {Artrith}}, \bibinfo {author} {\bibfnamefont {C.}~\bibnamefont {Ophus}}, \bibinfo {author} {\bibfnamefont {W.}~\bibnamefont {Huang}}, \bibinfo {author} {\bibfnamefont {Z.}~\bibnamefont {Cai}}, \bibinfo {author} {\bibfnamefont {T.}~\bibnamefont {Shi}}, \bibinfo {author} {\bibfnamefont {J.~C.}\ \bibnamefont {Kim}}, \bibinfo {author} {\bibfnamefont {H.}~\bibnamefont {Kim}},\ and\ \bibinfo {author} {\bibfnamefont {G.}~\bibnamefont {Ceder}},\ }\href {https://doi.org/10.1038/s41467-019-08490-w} {\bibfield  {journal} {\bibinfo  {journal} {Nature Communications}\ }\textbf {\bibinfo {volume} {10}},\ \bibinfo {pages} {592} (\bibinfo {year} {2019})}\BibitemShut {NoStop}%
\bibitem [{\citenamefont {Cao}\ \emph {et~al.}(2020)\citenamefont {Cao}, \citenamefont {Chen}, \citenamefont {Jin}, \citenamefont {Liu},\ and\ \citenamefont {Li}}]{Cao2020ACSAMI}%
  \BibitemOpen
  \bibfield  {author} {\bibinfo {author} {\bibfnamefont {B.}~\bibnamefont {Cao}}, \bibinfo {author} {\bibfnamefont {S.}~\bibnamefont {Chen}}, \bibinfo {author} {\bibfnamefont {X.}~\bibnamefont {Jin}}, \bibinfo {author} {\bibfnamefont {J.}~\bibnamefont {Liu}},\ and\ \bibinfo {author} {\bibfnamefont {T.}~\bibnamefont {Li}},\ }\href {https://doi.org/10.1021/acsami.0c18483} {\bibfield  {journal} {\bibinfo  {journal} {ACS Applied Materials \& Interfaces}\ }\textbf {\bibinfo {volume} {12}},\ \bibinfo {pages} {57245} (\bibinfo {year} {2020})}\BibitemShut {NoStop}%
\bibitem [{\citenamefont {Zhang}\ \emph {et~al.}(2020)\citenamefont {Zhang}, \citenamefont {Zhao}, \citenamefont {Ding}, \citenamefont {Chong}, \citenamefont {Jia}, \citenamefont {Ophus}, \citenamefont {Asta}, \citenamefont {Ritchie},\ and\ \citenamefont {Minor}}]{Zhang_Natue_2020_Short}%
  \BibitemOpen
  \bibfield  {author} {\bibinfo {author} {\bibfnamefont {R.}~\bibnamefont {Zhang}}, \bibinfo {author} {\bibfnamefont {S.}~\bibnamefont {Zhao}}, \bibinfo {author} {\bibfnamefont {J.}~\bibnamefont {Ding}}, \bibinfo {author} {\bibfnamefont {Y.}~\bibnamefont {Chong}}, \bibinfo {author} {\bibfnamefont {T.}~\bibnamefont {Jia}}, \bibinfo {author} {\bibfnamefont {C.}~\bibnamefont {Ophus}}, \bibinfo {author} {\bibfnamefont {M.}~\bibnamefont {Asta}}, \bibinfo {author} {\bibfnamefont {R.~O.}\ \bibnamefont {Ritchie}},\ and\ \bibinfo {author} {\bibfnamefont {A.~M.}\ \bibnamefont {Minor}},\ }\href {https://doi.org/10.1038/s41586-020-2275-z} {\bibfield  {journal} {\bibinfo  {journal} {Nature}\ }\textbf {\bibinfo {volume} {581}},\ \bibinfo {pages} {283} (\bibinfo {year} {2020})}\BibitemShut {NoStop}%
\bibitem [{\citenamefont {Roychowdhury}\ \emph {et~al.}(2021)\citenamefont {Roychowdhury}, \citenamefont {Ghosh}, \citenamefont {Arora}, \citenamefont {Samanta}, \citenamefont {Xie}, \citenamefont {Singh}, \citenamefont {Soni}, \citenamefont {He}, \citenamefont {Waghmare},\ and\ \citenamefont {Biswas}}]{Roychowdhury_science_2021_enhanced}%
  \BibitemOpen
  \bibfield  {author} {\bibinfo {author} {\bibfnamefont {S.}~\bibnamefont {Roychowdhury}}, \bibinfo {author} {\bibfnamefont {T.}~\bibnamefont {Ghosh}}, \bibinfo {author} {\bibfnamefont {R.}~\bibnamefont {Arora}}, \bibinfo {author} {\bibfnamefont {M.}~\bibnamefont {Samanta}}, \bibinfo {author} {\bibfnamefont {L.}~\bibnamefont {Xie}}, \bibinfo {author} {\bibfnamefont {N.~K.}\ \bibnamefont {Singh}}, \bibinfo {author} {\bibfnamefont {A.}~\bibnamefont {Soni}}, \bibinfo {author} {\bibfnamefont {J.}~\bibnamefont {He}}, \bibinfo {author} {\bibfnamefont {U.~V.}\ \bibnamefont {Waghmare}},\ and\ \bibinfo {author} {\bibfnamefont {K.}~\bibnamefont {Biswas}},\ }\href {https://doi.org/10.1126/science.abb3517} {\bibfield  {journal} {\bibinfo  {journal} {Science}\ }\textbf {\bibinfo {volume} {371}},\ \bibinfo {pages} {722} (\bibinfo {year} {2021})}\BibitemShut {NoStop}%
\bibitem [{\citenamefont {Jiang}\ \emph {et~al.}(2021)\citenamefont {Jiang}, \citenamefont {Yu}, \citenamefont {Cui}, \citenamefont {Liu}, \citenamefont {Xie}, \citenamefont {Liao}, \citenamefont {Zhang}, \citenamefont {Huang}, \citenamefont {Ning}, \citenamefont {Jia}, \citenamefont {Zhu}, \citenamefont {Bai}, \citenamefont {Chen}, \citenamefont {Pennycook},\ and\ \citenamefont {He}}]{Jiang_Science_2021_high}%
  \BibitemOpen
  \bibfield  {author} {\bibinfo {author} {\bibfnamefont {B.}~\bibnamefont {Jiang}}, \bibinfo {author} {\bibfnamefont {Y.}~\bibnamefont {Yu}}, \bibinfo {author} {\bibfnamefont {J.}~\bibnamefont {Cui}}, \bibinfo {author} {\bibfnamefont {X.}~\bibnamefont {Liu}}, \bibinfo {author} {\bibfnamefont {L.}~\bibnamefont {Xie}}, \bibinfo {author} {\bibfnamefont {J.}~\bibnamefont {Liao}}, \bibinfo {author} {\bibfnamefont {Q.}~\bibnamefont {Zhang}}, \bibinfo {author} {\bibfnamefont {Y.}~\bibnamefont {Huang}}, \bibinfo {author} {\bibfnamefont {S.}~\bibnamefont {Ning}}, \bibinfo {author} {\bibfnamefont {B.}~\bibnamefont {Jia}}, \bibinfo {author} {\bibfnamefont {B.}~\bibnamefont {Zhu}}, \bibinfo {author} {\bibfnamefont {S.}~\bibnamefont {Bai}}, \bibinfo {author} {\bibfnamefont {L.}~\bibnamefont {Chen}}, \bibinfo {author} {\bibfnamefont {S.~J.}\ \bibnamefont {Pennycook}},\ and\ \bibinfo {author} {\bibfnamefont {J.}~\bibnamefont {He}},\ }\href {https://doi.org/10.1126/science.abe1292} {\bibfield  {journal} {\bibinfo  {journal}
  {Science}\ }\textbf {\bibinfo {volume} {371}},\ \bibinfo {pages} {830} (\bibinfo {year} {2021})}\BibitemShut {NoStop}%
\bibitem [{\citenamefont {Jin}\ \emph {et~al.}(2021)\citenamefont {Jin}, \citenamefont {Chen},\ and\ \citenamefont {Li}}]{jin2021short}%
  \BibitemOpen
  \bibfield  {author} {\bibinfo {author} {\bibfnamefont {X.}~\bibnamefont {Jin}}, \bibinfo {author} {\bibfnamefont {S.}~\bibnamefont {Chen}},\ and\ \bibinfo {author} {\bibfnamefont {T.}~\bibnamefont {Li}},\ }\href@noop {} {\bibfield  {journal} {\bibinfo  {journal} {Physical Review Materials}\ }\textbf {\bibinfo {volume} {5}},\ \bibinfo {pages} {104606} (\bibinfo {year} {2021})}\BibitemShut {NoStop}%
\bibitem [{\citenamefont {Jin}\ \emph {et~al.}(2022)\citenamefont {Jin}, \citenamefont {Chen},\ and\ \citenamefont {Li}}]{jin2022coexistence}%
  \BibitemOpen
  \bibfield  {author} {\bibinfo {author} {\bibfnamefont {X.}~\bibnamefont {Jin}}, \bibinfo {author} {\bibfnamefont {S.}~\bibnamefont {Chen}},\ and\ \bibinfo {author} {\bibfnamefont {T.}~\bibnamefont {Li}},\ }\href {https://doi.org/10.1038/s43246-022-00289-5} {\bibfield  {journal} {\bibinfo  {journal} {Communications Materials}\ }\textbf {\bibinfo {volume} {3}},\ \bibinfo {pages} {66} (\bibinfo {year} {2022})}\BibitemShut {NoStop}%
\bibitem [{\citenamefont {Chen}\ and\ \citenamefont {Li}(2023)}]{chen2023impacts}%
  \BibitemOpen
  \bibfield  {author} {\bibinfo {author} {\bibfnamefont {S.}~\bibnamefont {Chen}}\ and\ \bibinfo {author} {\bibfnamefont {T.}~\bibnamefont {Li}},\ }in\ \href@noop {} {\emph {\bibinfo {booktitle} {APS March Meeting Abstracts}}},\ Vol.\ \bibinfo {volume} {2023}\ (\bibinfo {year} {2023})\ pp.\ \bibinfo {pages} {F61--008}\BibitemShut {NoStop}%
\bibitem [{\citenamefont {Corley-Wiciak}\ \emph {et~al.}(2023)\citenamefont {Corley-Wiciak}, \citenamefont {Chen}, \citenamefont {Concepci\'on}, \citenamefont {Zoellner}, \citenamefont {Gr\"utzmacher}, \citenamefont {Buca}, \citenamefont {Li}, \citenamefont {Capellini},\ and\ \citenamefont {Spirito}}]{Corley-Wiciak2023PRApplied}%
  \BibitemOpen
  \bibfield  {author} {\bibinfo {author} {\bibfnamefont {A.~A.}\ \bibnamefont {Corley-Wiciak}}, \bibinfo {author} {\bibfnamefont {S.}~\bibnamefont {Chen}}, \bibinfo {author} {\bibfnamefont {O.}~\bibnamefont {Concepci\'on}}, \bibinfo {author} {\bibfnamefont {M.~H.}\ \bibnamefont {Zoellner}}, \bibinfo {author} {\bibfnamefont {D.}~\bibnamefont {Gr\"utzmacher}}, \bibinfo {author} {\bibfnamefont {D.}~\bibnamefont {Buca}}, \bibinfo {author} {\bibfnamefont {T.}~\bibnamefont {Li}}, \bibinfo {author} {\bibfnamefont {G.}~\bibnamefont {Capellini}},\ and\ \bibinfo {author} {\bibfnamefont {D.}~\bibnamefont {Spirito}},\ }\href {https://doi.org/10.1103/PhysRevApplied.20.024021} {\bibfield  {journal} {\bibinfo  {journal} {Phys. Rev. Appl.}\ }\textbf {\bibinfo {volume} {20}},\ \bibinfo {pages} {024021} (\bibinfo {year} {2023})}\BibitemShut {NoStop}%
\bibitem [{\citenamefont {Jin}\ \emph {et~al.}(2023)\citenamefont {Jin}, \citenamefont {Chen}, \citenamefont {Lemkan},\ and\ \citenamefont {Li}}]{JinChenLiPRM2023}%
  \BibitemOpen
  \bibfield  {author} {\bibinfo {author} {\bibfnamefont {X.}~\bibnamefont {Jin}}, \bibinfo {author} {\bibfnamefont {S.}~\bibnamefont {Chen}}, \bibinfo {author} {\bibfnamefont {C.}~\bibnamefont {Lemkan}},\ and\ \bibinfo {author} {\bibfnamefont {T.}~\bibnamefont {Li}},\ }\href {https://doi.org/10.1103/PhysRevMaterials.7.L111601} {\bibfield  {journal} {\bibinfo  {journal} {Phys. Rev. Mater.}\ }\textbf {\bibinfo {volume} {7}},\ \bibinfo {pages} {L111601} (\bibinfo {year} {2023})}\BibitemShut {NoStop}%
\bibitem [{\citenamefont {Chen}\ \emph {et~al.}(2024)\citenamefont {Chen}, \citenamefont {Jin}, \citenamefont {Zhao},\ and\ \citenamefont {Li}}]{chen2024intricate}%
  \BibitemOpen
  \bibfield  {author} {\bibinfo {author} {\bibfnamefont {S.}~\bibnamefont {Chen}}, \bibinfo {author} {\bibfnamefont {X.}~\bibnamefont {Jin}}, \bibinfo {author} {\bibfnamefont {W.}~\bibnamefont {Zhao}},\ and\ \bibinfo {author} {\bibfnamefont {T.}~\bibnamefont {Li}},\ }\href {https://doi.org/10.1103/PhysRevMaterials.8.043805} {\bibfield  {journal} {\bibinfo  {journal} {Physical Review Materials}\ }\textbf {\bibinfo {volume} {8}},\ \bibinfo {pages} {043805} (\bibinfo {year} {2024})}\BibitemShut {NoStop}%
\bibitem [{\citenamefont {Liang}\ \emph {et~al.}(2024)\citenamefont {Liang}, \citenamefont {Chen}, \citenamefont {Jin}, \citenamefont {West}, \citenamefont {Yu}, \citenamefont {Li},\ and\ \citenamefont {Zhang}}]{liang2024group}%
  \BibitemOpen
  \bibfield  {author} {\bibinfo {author} {\bibfnamefont {Y.}~\bibnamefont {Liang}}, \bibinfo {author} {\bibfnamefont {S.}~\bibnamefont {Chen}}, \bibinfo {author} {\bibfnamefont {X.}~\bibnamefont {Jin}}, \bibinfo {author} {\bibfnamefont {D.}~\bibnamefont {West}}, \bibinfo {author} {\bibfnamefont {S.-Q.}\ \bibnamefont {Yu}}, \bibinfo {author} {\bibfnamefont {T.}~\bibnamefont {Li}},\ and\ \bibinfo {author} {\bibfnamefont {S.}~\bibnamefont {Zhang}},\ }\href {https://doi.org/10.1038/s41524-024-01271-0} {\bibfield  {journal} {\bibinfo  {journal} {npj Computational Materials}\ }\textbf {\bibinfo {volume} {10}},\ \bibinfo {pages} {82} (\bibinfo {year} {2024})}\BibitemShut {NoStop}%
\bibitem [{\citenamefont {Liang}\ \emph {et~al.}(2025)\citenamefont {Liang}, \citenamefont {West}, \citenamefont {Chen}, \citenamefont {Liu}, \citenamefont {Li},\ and\ \citenamefont {Zhang}}]{LIANG2025MT}%
  \BibitemOpen
  \bibfield  {author} {\bibinfo {author} {\bibfnamefont {Y.}~\bibnamefont {Liang}}, \bibinfo {author} {\bibfnamefont {D.}~\bibnamefont {West}}, \bibinfo {author} {\bibfnamefont {S.}~\bibnamefont {Chen}}, \bibinfo {author} {\bibfnamefont {J.}~\bibnamefont {Liu}}, \bibinfo {author} {\bibfnamefont {T.}~\bibnamefont {Li}},\ and\ \bibinfo {author} {\bibfnamefont {S.}~\bibnamefont {Zhang}},\ }\href {https://doi.org/https://doi.org/10.1016/j.mattod.2025.03.017} {\bibfield  {journal} {\bibinfo  {journal} {Materials Today}\ }\textbf {\bibinfo {volume} {86}},\ \bibinfo {pages} {115} (\bibinfo {year} {2025})}\BibitemShut {NoStop}%
\bibitem [{\citenamefont {Vogl}\ \emph {et~al.}(2025)\citenamefont {Vogl}, \citenamefont {Chen}, \citenamefont {Schweizer}, \citenamefont {Jin}, \citenamefont {Yu}, \citenamefont {Liu}, \citenamefont {Li},\ and\ \citenamefont {Minor}}]{vogl2025identification}%
  \BibitemOpen
  \bibfield  {author} {\bibinfo {author} {\bibfnamefont {L.~M.}\ \bibnamefont {Vogl}}, \bibinfo {author} {\bibfnamefont {S.}~\bibnamefont {Chen}}, \bibinfo {author} {\bibfnamefont {P.}~\bibnamefont {Schweizer}}, \bibinfo {author} {\bibfnamefont {X.}~\bibnamefont {Jin}}, \bibinfo {author} {\bibfnamefont {S.-Q.}\ \bibnamefont {Yu}}, \bibinfo {author} {\bibfnamefont {J.}~\bibnamefont {Liu}}, \bibinfo {author} {\bibfnamefont {T.}~\bibnamefont {Li}},\ and\ \bibinfo {author} {\bibfnamefont {A.~M.}\ \bibnamefont {Minor}},\ }\href {https://doi.org/10.1126/science.adu0719} {\bibfield  {journal} {\bibinfo  {journal} {Science}\ }\textbf {\bibinfo {volume} {389}},\ \bibinfo {pages} {1342} (\bibinfo {year} {2025})}\BibitemShut {NoStop}%
\bibitem [{\citenamefont {Jin}\ \emph {et~al.}(2025)\citenamefont {Jin}, \citenamefont {Chen},\ and\ \citenamefont {Li}}]{Jin2025IEEE}%
  \BibitemOpen
  \bibfield  {author} {\bibinfo {author} {\bibfnamefont {X.}~\bibnamefont {Jin}}, \bibinfo {author} {\bibfnamefont {S.}~\bibnamefont {Chen}},\ and\ \bibinfo {author} {\bibfnamefont {T.}~\bibnamefont {Li}},\ }\href {https://doi.org/10.1109/JSTQE.2024.3419713} {\bibfield  {journal} {\bibinfo  {journal} {IEEE Journal of Selected Topics in Quantum Electronics}\ }\textbf {\bibinfo {volume} {31}},\ \bibinfo {pages} {1} (\bibinfo {year} {2025})}\BibitemShut {NoStop}%
\bibitem [{\citenamefont {Liu}\ \emph {et~al.}(2026)\citenamefont {Liu}, \citenamefont {Liang}, \citenamefont {Eldose}, \citenamefont {Chen}, \citenamefont {Jin}, \citenamefont {Zhao}, \citenamefont {Shah}, \citenamefont {Bae}, \citenamefont {Concepcion}, \citenamefont {de~Oliveira}, \citenamefont {Bikmukhametov}, \citenamefont {Wang}, \citenamefont {Zeng}, \citenamefont {Buca}, \citenamefont {Mortazavi}, \citenamefont {West}, \citenamefont {Zhang}, \citenamefont {Li}, \citenamefont {Salamo}, \citenamefont {Yu},\ and\ \citenamefont {Liu}}]{liu2026_APT-GeSn-CVD-MBE}%
  \BibitemOpen
  \bibfield  {author} {\bibinfo {author} {\bibfnamefont {S.}~\bibnamefont {Liu}}, \bibinfo {author} {\bibfnamefont {Y.}~\bibnamefont {Liang}}, \bibinfo {author} {\bibfnamefont {N.~M.}\ \bibnamefont {Eldose}}, \bibinfo {author} {\bibfnamefont {S.}~\bibnamefont {Chen}}, \bibinfo {author} {\bibfnamefont {X.}~\bibnamefont {Jin}}, \bibinfo {author} {\bibfnamefont {H.}~\bibnamefont {Zhao}}, \bibinfo {author} {\bibfnamefont {M.}~\bibnamefont {Shah}}, \bibinfo {author} {\bibfnamefont {J.-H.}\ \bibnamefont {Bae}}, \bibinfo {author} {\bibfnamefont {O.}~\bibnamefont {Concepcion}}, \bibinfo {author} {\bibfnamefont {F.~M.}\ \bibnamefont {de~Oliveira}}, \bibinfo {author} {\bibfnamefont {I.}~\bibnamefont {Bikmukhametov}}, \bibinfo {author} {\bibfnamefont {X.}~\bibnamefont {Wang}}, \bibinfo {author} {\bibfnamefont {Y.}~\bibnamefont {Zeng}}, \bibinfo {author} {\bibfnamefont {D.}~\bibnamefont {Buca}}, \bibinfo {author} {\bibfnamefont {M.}~\bibnamefont {Mortazavi}}, \bibinfo {author} {\bibfnamefont {D.}~\bibnamefont {West}},
  \bibinfo {author} {\bibfnamefont {S.}~\bibnamefont {Zhang}}, \bibinfo {author} {\bibfnamefont {T.}~\bibnamefont {Li}}, \bibinfo {author} {\bibfnamefont {G.~J.}\ \bibnamefont {Salamo}}, \bibinfo {author} {\bibfnamefont {S.-Q.}\ \bibnamefont {Yu}},\ and\ \bibinfo {author} {\bibfnamefont {J.}~\bibnamefont {Liu}},\ }\href {https://arxiv.org/abs/2407.02767} {\bibinfo {title} {Atomic short-range order control of {GeSn} as a new degree of freedom for band engineering}} (\bibinfo {year} {2026}),\ \Eprint {https://arxiv.org/abs/2407.02767} {arXiv:2407.02767} \BibitemShut {NoStop}%
\bibitem [{\citenamefont {Attiaoui}\ \emph {et~al.}(2026)\citenamefont {Attiaoui}, \citenamefont {Chen}, \citenamefont {Woicik}, \citenamefont {Lentz}, \citenamefont {Vogl}, \citenamefont {Meyer}, \citenamefont {Mukherjee}, \citenamefont {Minor}, \citenamefont {Li},\ and\ \citenamefont {McIntyre}}]{Anis2026shining}%
  \BibitemOpen
  \bibfield  {author} {\bibinfo {author} {\bibfnamefont {A.}~\bibnamefont {Attiaoui}}, \bibinfo {author} {\bibfnamefont {S.}~\bibnamefont {Chen}}, \bibinfo {author} {\bibfnamefont {J.~C.}\ \bibnamefont {Woicik}}, \bibinfo {author} {\bibfnamefont {J.~Z.}\ \bibnamefont {Lentz}}, \bibinfo {author} {\bibfnamefont {L.~M.}\ \bibnamefont {Vogl}}, \bibinfo {author} {\bibfnamefont {J.~E.}\ \bibnamefont {Meyer}}, \bibinfo {author} {\bibfnamefont {K.}~\bibnamefont {Mukherjee}}, \bibinfo {author} {\bibfnamefont {A.}~\bibnamefont {Minor}}, \bibinfo {author} {\bibfnamefont {T.}~\bibnamefont {Li}},\ and\ \bibinfo {author} {\bibfnamefont {P.~C.}\ \bibnamefont {McIntyre}},\ }\href {https://arxiv.org/abs/2603.27876} {\bibinfo {title} {Shining light on short-range atomic ordering in semiconductors alloys}} (\bibinfo {year} {2026}),\ \Eprint {https://arxiv.org/abs/2603.27876} {arXiv:2603.27876} \BibitemShut {NoStop}%
\bibitem [{\citenamefont {Bhattarai}\ \emph {et~al.}(2024)\citenamefont {Bhattarai}, \citenamefont {Zhang}, \citenamefont {Xu}, \citenamefont {Gu}, \citenamefont {Meng},\ and\ \citenamefont {Meng}}]{bhattarai_effect_2024}%
  \BibitemOpen
  \bibfield  {author} {\bibinfo {author} {\bibfnamefont {B.}~\bibnamefont {Bhattarai}}, \bibinfo {author} {\bibfnamefont {X.}~\bibnamefont {Zhang}}, \bibinfo {author} {\bibfnamefont {W.}~\bibnamefont {Xu}}, \bibinfo {author} {\bibfnamefont {Y.}~\bibnamefont {Gu}}, \bibinfo {author} {\bibfnamefont {W.~J.}\ \bibnamefont {Meng}},\ and\ \bibinfo {author} {\bibfnamefont {A.~C.}\ \bibnamefont {Meng}},\ }\href {https://doi.org/10.1039/D4MH00551A} {\bibfield  {journal} {\bibinfo  {journal} {Materials Horizons}\ }\textbf {\bibinfo {volume} {11}},\ \bibinfo {pages} {5402} (\bibinfo {year} {2024})}\BibitemShut {NoStop}%
\bibitem [{\citenamefont {Hart}\ \emph {et~al.}(2025)\citenamefont {Hart}, \citenamefont {Lang}, \citenamefont {Hardy}, \citenamefont {Mukhopadhyay}, \citenamefont {Gokhale}, \citenamefont {Champlain}, \citenamefont {Hudak}, \citenamefont {Giribaldi}, \citenamefont {Colombo}, \citenamefont {Rinaldi},\ and\ \citenamefont {Downey}}]{hart2025enhanced}%
  \BibitemOpen
  \bibfield  {author} {\bibinfo {author} {\bibfnamefont {J.~L.}\ \bibnamefont {Hart}}, \bibinfo {author} {\bibfnamefont {A.~C.}\ \bibnamefont {Lang}}, \bibinfo {author} {\bibfnamefont {M.~T.}\ \bibnamefont {Hardy}}, \bibinfo {author} {\bibfnamefont {S.}~\bibnamefont {Mukhopadhyay}}, \bibinfo {author} {\bibfnamefont {V.~J.}\ \bibnamefont {Gokhale}}, \bibinfo {author} {\bibfnamefont {J.~G.}\ \bibnamefont {Champlain}}, \bibinfo {author} {\bibfnamefont {B.~M.}\ \bibnamefont {Hudak}}, \bibinfo {author} {\bibfnamefont {G.}~\bibnamefont {Giribaldi}}, \bibinfo {author} {\bibfnamefont {L.}~\bibnamefont {Colombo}}, \bibinfo {author} {\bibfnamefont {M.}~\bibnamefont {Rinaldi}},\ and\ \bibinfo {author} {\bibfnamefont {B.~P.}\ \bibnamefont {Downey}},\ }\href {https://arxiv.org/abs/2512.19599} {\bibinfo {title} {Enhanced permittivity in wurtzite {ScAlN} through nanoscale {Sc} clustering}} (\bibinfo {year} {2025}),\ \Eprint {https://arxiv.org/abs/2512.19599} {arXiv:2512.19599} \BibitemShut {NoStop}%
\bibitem [{\citenamefont {Das}\ \emph {et~al.}(2026)\citenamefont {Das}, \citenamefont {Nguyen}, \citenamefont {Savant}, \citenamefont {Gothandapani}, \citenamefont {Xing}, \citenamefont {Jena},\ and\ \citenamefont {Mazumder}}]{Das2026-APL_APT_ScAlN}%
  \BibitemOpen
  \bibfield  {author} {\bibinfo {author} {\bibfnamefont {S.}~\bibnamefont {Das}}, \bibinfo {author} {\bibfnamefont {T.-S.}\ \bibnamefont {Nguyen}}, \bibinfo {author} {\bibfnamefont {C.}~\bibnamefont {Savant}}, \bibinfo {author} {\bibfnamefont {K.}~\bibnamefont {Gothandapani}}, \bibinfo {author} {\bibfnamefont {H.~G.}\ \bibnamefont {Xing}}, \bibinfo {author} {\bibfnamefont {D.}~\bibnamefont {Jena}},\ and\ \bibinfo {author} {\bibfnamefont {B.}~\bibnamefont {Mazumder}},\ }\href {https://doi.org/10.1063/5.0325478} {\bibfield  {journal} {\bibinfo  {journal} {Applied Physics Letters}\ }\textbf {\bibinfo {volume} {128}},\ \bibinfo {pages} {142905} (\bibinfo {year} {2026})}\BibitemShut {NoStop}%
\bibitem [{\citenamefont {Akiyama}\ \emph {et~al.}(2009)\citenamefont {Akiyama}, \citenamefont {Kamohara}, \citenamefont {Kano}, \citenamefont {Teshigahara}, \citenamefont {Takeuchi},\ and\ \citenamefont {Kawahara}}]{Akiyama2009AMenhancement}%
  \BibitemOpen
  \bibfield  {author} {\bibinfo {author} {\bibfnamefont {M.}~\bibnamefont {Akiyama}}, \bibinfo {author} {\bibfnamefont {T.}~\bibnamefont {Kamohara}}, \bibinfo {author} {\bibfnamefont {K.}~\bibnamefont {Kano}}, \bibinfo {author} {\bibfnamefont {A.}~\bibnamefont {Teshigahara}}, \bibinfo {author} {\bibfnamefont {Y.}~\bibnamefont {Takeuchi}},\ and\ \bibinfo {author} {\bibfnamefont {N.}~\bibnamefont {Kawahara}},\ }\href {https://doi.org/https://doi.org/10.1002/adma.200802611} {\bibfield  {journal} {\bibinfo  {journal} {Advanced Materials}\ }\textbf {\bibinfo {volume} {21}},\ \bibinfo {pages} {593} (\bibinfo {year} {2009})}\BibitemShut {NoStop}%
\bibitem [{\citenamefont {Huang}\ \emph {et~al.}(2026)\citenamefont {Huang}, \citenamefont {Li}, \citenamefont {Guo}, \citenamefont {Wen}, \citenamefont {Srolovitz}, \citenamefont {Chen}, \citenamefont {Chen},\ and\ \citenamefont {Liu}}]{Huang_PRL_2026_atomistic}%
  \BibitemOpen
  \bibfield  {author} {\bibinfo {author} {\bibfnamefont {J.}~\bibnamefont {Huang}}, \bibinfo {author} {\bibfnamefont {J.}~\bibnamefont {Li}}, \bibinfo {author} {\bibfnamefont {X.}~\bibnamefont {Guo}}, \bibinfo {author} {\bibfnamefont {T.}~\bibnamefont {Wen}}, \bibinfo {author} {\bibfnamefont {D.~J.}\ \bibnamefont {Srolovitz}}, \bibinfo {author} {\bibfnamefont {Z.}~\bibnamefont {Chen}}, \bibinfo {author} {\bibfnamefont {Z.}~\bibnamefont {Chen}},\ and\ \bibinfo {author} {\bibfnamefont {S.}~\bibnamefont {Liu}},\ }\href {https://doi.org/10.1103/lfdh-86x6} {\bibfield  {journal} {\bibinfo  {journal} {Phys. Rev. Lett.}\ }\textbf {\bibinfo {volume} {136}},\ \bibinfo {pages} {026801} (\bibinfo {year} {2026})}\BibitemShut {NoStop}%
\bibitem [{\citenamefont {Zheng}\ \emph {et~al.}(2026)\citenamefont {Zheng}, \citenamefont {Paillard}, \citenamefont {Wang}, \citenamefont {Chen}, \citenamefont {Zhao}, \citenamefont {Xie},\ and\ \citenamefont {Bellaiche}}]{Zheng_PRL_2026_Domain}%
  \BibitemOpen
  \bibfield  {author} {\bibinfo {author} {\bibfnamefont {X.}~\bibnamefont {Zheng}}, \bibinfo {author} {\bibfnamefont {C.}~\bibnamefont {Paillard}}, \bibinfo {author} {\bibfnamefont {D.}~\bibnamefont {Wang}}, \bibinfo {author} {\bibfnamefont {P.}~\bibnamefont {Chen}}, \bibinfo {author} {\bibfnamefont {H.~J.}\ \bibnamefont {Zhao}}, \bibinfo {author} {\bibfnamefont {Y.}~\bibnamefont {Xie}},\ and\ \bibinfo {author} {\bibfnamefont {L.}~\bibnamefont {Bellaiche}},\ }\href {https://doi.org/10.1103/s8qs-nnzg} {\bibfield  {journal} {\bibinfo  {journal} {Phys. Rev. Lett.}\ }\textbf {\bibinfo {volume} {136}},\ \bibinfo {pages} {206102} (\bibinfo {year} {2026})}\BibitemShut {NoStop}%
\bibitem [{\citenamefont {Vogl}\ \emph {et~al.}(2024)\citenamefont {Vogl}, \citenamefont {Schweizer}, \citenamefont {Chen}, \citenamefont {Jin}, \citenamefont {Yu}, \citenamefont {Byrne}, \citenamefont {Allen}, \citenamefont {Liu}, \citenamefont {Li},\ and\ \citenamefont {Minor}}]{Vogl_MM_2024_exploring}%
  \BibitemOpen
  \bibfield  {author} {\bibinfo {author} {\bibfnamefont {L.~M.}\ \bibnamefont {Vogl}}, \bibinfo {author} {\bibfnamefont {P.}~\bibnamefont {Schweizer}}, \bibinfo {author} {\bibfnamefont {S.}~\bibnamefont {Chen}}, \bibinfo {author} {\bibfnamefont {X.}~\bibnamefont {Jin}}, \bibinfo {author} {\bibfnamefont {S.-Q.}\ \bibnamefont {Yu}}, \bibinfo {author} {\bibfnamefont {D.~O.}\ \bibnamefont {Byrne}}, \bibinfo {author} {\bibfnamefont {F.~I.}\ \bibnamefont {Allen}}, \bibinfo {author} {\bibfnamefont {J.}~\bibnamefont {Liu}}, \bibinfo {author} {\bibfnamefont {T.}~\bibnamefont {Li}},\ and\ \bibinfo {author} {\bibfnamefont {A.~M.}\ \bibnamefont {Minor}},\ }\href {https://doi.org/10.1093/mam/ozae044.565} {\bibfield  {journal} {\bibinfo  {journal} {Microscopy and Microanalysis}\ }\textbf {\bibinfo {volume} {30}},\ \bibinfo {pages} {ozae044.565} (\bibinfo {year} {2024})}\BibitemShut {NoStop}%
\end{thebibliography}%


\begin{thebibliography}{15}%
\makeatletter
\providecommand \@ifxundefined [1]{%
 \@ifx{#1\undefined}
}%
\providecommand \@ifnum [1]{%
 \ifnum #1\expandafter \@firstoftwo
 \else \expandafter \@secondoftwo
 \fi
}%
\providecommand \@ifx [1]{%
 \ifx #1\expandafter \@firstoftwo
 \else \expandafter \@secondoftwo
 \fi
}%
\providecommand \natexlab [1]{#1}%
\providecommand \enquote  [1]{``#1''}%
\providecommand \bibnamefont  [1]{#1}%
\providecommand \bibfnamefont [1]{#1}%
\providecommand \citenamefont [1]{#1}%
\providecommand \href@noop [0]{\@secondoftwo}%
\providecommand \href [0]{\begingroup \@sanitize@url \@href}%
\providecommand \@href[1]{\@@startlink{#1}\@@href}%
\providecommand \@@href[1]{\endgroup#1\@@endlink}%
\providecommand \@sanitize@url [0]{\catcode `\\12\catcode `\$12\catcode `\&12\catcode `\#12\catcode `\^12\catcode `\_12\catcode `\%12\relax}%
\providecommand \@@startlink[1]{}%
\providecommand \@@endlink[0]{}%
\providecommand \url  [0]{\begingroup\@sanitize@url \@url }%
\providecommand \@url [1]{\endgroup\@href {#1}{\urlprefix }}%
\providecommand \urlprefix  [0]{URL }%
\providecommand \Eprint [0]{\href }%
\providecommand \doibase [0]{https://doi.org/}%
\providecommand \selectlanguage [0]{\@gobble}%
\providecommand \bibinfo  [0]{\@secondoftwo}%
\providecommand \bibfield  [0]{\@secondoftwo}%
\providecommand \translation [1]{[#1]}%
\providecommand \BibitemOpen [0]{}%
\providecommand \bibitemStop [0]{}%
\providecommand \bibitemNoStop [0]{.\EOS\space}%
\providecommand \EOS [0]{\spacefactor3000\relax}%
\providecommand \BibitemShut  [1]{\csname bibitem#1\endcsname}%
\let\auto@bib@innerbib\@empty
\bibitem [{\citenamefont {Kresse}\ and\ \citenamefont {Joubert}(1999)}]{Kresse:1999tq}%
  \BibitemOpen
  \bibfield  {author} {\bibinfo {author} {\bibfnamefont {G.}~\bibnamefont {Kresse}}\ and\ \bibinfo {author} {\bibfnamefont {D.}~\bibnamefont {Joubert}},\ }\href {https://doi.org/10.1103/PhysRevB.59.1758} {\bibfield  {journal} {\bibinfo  {journal} {Phys. Rev. B}\ }\textbf {\bibinfo {volume} {59}},\ \bibinfo {pages} {1758} (\bibinfo {year} {1999})}\BibitemShut {NoStop}%
\bibitem [{\citenamefont {Kresse}\ and\ \citenamefont {Furthmüller}(1996{\natexlab{a}})}]{Kresse:1996kg}%
  \BibitemOpen
  \bibfield  {author} {\bibinfo {author} {\bibfnamefont {G.}~\bibnamefont {Kresse}}\ and\ \bibinfo {author} {\bibfnamefont {J.}~\bibnamefont {Furthmüller}},\ }\href {https://doi.org/10.1016/0927-0256(96)00008-0} {\bibfield  {journal} {\bibinfo  {journal} {Computational Materials Science}\ }\textbf {\bibinfo {volume} {6}},\ \bibinfo {pages} {15} (\bibinfo {year} {1996}{\natexlab{a}})}\BibitemShut {NoStop}%
\bibitem [{\citenamefont {Kresse}\ and\ \citenamefont {Furthmüller}(1996{\natexlab{b}})}]{Kresse:1996vf}%
  \BibitemOpen
  \bibfield  {author} {\bibinfo {author} {\bibfnamefont {G.}~\bibnamefont {Kresse}}\ and\ \bibinfo {author} {\bibfnamefont {J.}~\bibnamefont {Furthmüller}},\ }\href {https://doi.org/10.1103/PhysRevB.54.11169} {\bibfield  {journal} {\bibinfo  {journal} {Phys. Rev. B}\ }\textbf {\bibinfo {volume} {54}},\ \bibinfo {pages} {11169} (\bibinfo {year} {1996}{\natexlab{b}})}\BibitemShut {NoStop}%
\bibitem [{\citenamefont {Kresse}\ and\ \citenamefont {Hafner}(1993)}]{Kresse:1993ty}%
  \BibitemOpen
  \bibfield  {author} {\bibinfo {author} {\bibfnamefont {G.}~\bibnamefont {Kresse}}\ and\ \bibinfo {author} {\bibfnamefont {J.}~\bibnamefont {Hafner}},\ }\href {https://doi.org/10.1103/PhysRevB.47.558} {\bibfield  {journal} {\bibinfo  {journal} {Phys. Rev. B}\ }\textbf {\bibinfo {volume} {47}},\ \bibinfo {pages} {558} (\bibinfo {year} {1993})}\BibitemShut {NoStop}%
\bibitem [{\citenamefont {Perdew}\ \emph {et~al.}(1996)\citenamefont {Perdew}, \citenamefont {Burke},\ and\ \citenamefont {Ernzerhof}}]{PBE_GGA_PRL_1996}%
  \BibitemOpen
  \bibfield  {author} {\bibinfo {author} {\bibfnamefont {J.~P.}\ \bibnamefont {Perdew}}, \bibinfo {author} {\bibfnamefont {K.}~\bibnamefont {Burke}},\ and\ \bibinfo {author} {\bibfnamefont {M.}~\bibnamefont {Ernzerhof}},\ }\href {https://doi.org/10.1103/PhysRevLett.77.3865} {\bibfield  {journal} {\bibinfo  {journal} {Phys. Rev. Lett.}\ }\textbf {\bibinfo {volume} {77}},\ \bibinfo {pages} {3865} (\bibinfo {year} {1996})}\BibitemShut {NoStop}%
\bibitem [{\citenamefont {Monkhorst}\ and\ \citenamefont {Pack}(1976)}]{Monkhorst:1976cv}%
  \BibitemOpen
  \bibfield  {author} {\bibinfo {author} {\bibfnamefont {H.~J.}\ \bibnamefont {Monkhorst}}\ and\ \bibinfo {author} {\bibfnamefont {J.~D.}\ \bibnamefont {Pack}},\ }\href@noop {} {\bibfield  {journal} {\bibinfo  {journal} {Physical Review B}\ }\textbf {\bibinfo {volume} {13}},\ \bibinfo {pages} {5188} (\bibinfo {year} {1976})}\BibitemShut {NoStop}%
\bibitem [{\citenamefont {Zunger}\ \emph {et~al.}(1990)\citenamefont {Zunger}, \citenamefont {Wei}, \citenamefont {Ferreira},\ and\ \citenamefont {Bernard}}]{Zunger_PRL_1990_SQS}%
  \BibitemOpen
  \bibfield  {author} {\bibinfo {author} {\bibfnamefont {A.}~\bibnamefont {Zunger}}, \bibinfo {author} {\bibfnamefont {S.-H.}\ \bibnamefont {Wei}}, \bibinfo {author} {\bibfnamefont {L.~G.}\ \bibnamefont {Ferreira}},\ and\ \bibinfo {author} {\bibfnamefont {J.~E.}\ \bibnamefont {Bernard}},\ }\href {https://doi.org/10.1103/PhysRevLett.65.353} {\bibfield  {journal} {\bibinfo  {journal} {Phys. Rev. Lett.}\ }\textbf {\bibinfo {volume} {65}},\ \bibinfo {pages} {353} (\bibinfo {year} {1990})}\BibitemShut {NoStop}%
\bibitem [{\citenamefont {{van de Walle}}(2009)}]{Axel_2009_ATAT}%
  \BibitemOpen
  \bibfield  {author} {\bibinfo {author} {\bibfnamefont {A.}~\bibnamefont {{van de Walle}}},\ }\href {https://doi.org/https://doi.org/10.1016/j.calphad.2008.12.005} {\bibfield  {journal} {\bibinfo  {journal} {Calphad}\ }\textbf {\bibinfo {volume} {33}},\ \bibinfo {pages} {266} (\bibinfo {year} {2009})}\BibitemShut {NoStop}%
\bibitem [{\citenamefont {{van de Walle}}\ \emph {et~al.}(2013)\citenamefont {{van de Walle}}, \citenamefont {Tiwary}, \citenamefont {{de Jong}}, \citenamefont {Olmsted}, \citenamefont {Asta}, \citenamefont {Dick}, \citenamefont {Shin}, \citenamefont {Wang}, \citenamefont {Chen},\ and\ \citenamefont {Liu}}]{VANDEWALLE_2013_mcsqs}%
  \BibitemOpen
  \bibfield  {author} {\bibinfo {author} {\bibfnamefont {A.}~\bibnamefont {{van de Walle}}}, \bibinfo {author} {\bibfnamefont {P.}~\bibnamefont {Tiwary}}, \bibinfo {author} {\bibfnamefont {M.}~\bibnamefont {{de Jong}}}, \bibinfo {author} {\bibfnamefont {D.}~\bibnamefont {Olmsted}}, \bibinfo {author} {\bibfnamefont {M.}~\bibnamefont {Asta}}, \bibinfo {author} {\bibfnamefont {A.}~\bibnamefont {Dick}}, \bibinfo {author} {\bibfnamefont {D.}~\bibnamefont {Shin}}, \bibinfo {author} {\bibfnamefont {Y.}~\bibnamefont {Wang}}, \bibinfo {author} {\bibfnamefont {L.-Q.}\ \bibnamefont {Chen}},\ and\ \bibinfo {author} {\bibfnamefont {Z.-K.}\ \bibnamefont {Liu}},\ }\href {https://doi.org/https://doi.org/10.1016/j.calphad.2013.06.006} {\bibfield  {journal} {\bibinfo  {journal} {Calphad}\ }\textbf {\bibinfo {volume} {42}},\ \bibinfo {pages} {13} (\bibinfo {year} {2013})}\BibitemShut {NoStop}%
\bibitem [{\citenamefont {Metropolis}\ \emph {et~al.}(1953)\citenamefont {Metropolis}, \citenamefont {Rosenbluth}, \citenamefont {Rosenbluth}, \citenamefont {Teller},\ and\ \citenamefont {Teller}}]{Metropolis:1953in}%
  \BibitemOpen
  \bibfield  {author} {\bibinfo {author} {\bibfnamefont {N.}~\bibnamefont {Metropolis}}, \bibinfo {author} {\bibfnamefont {A.~W.}\ \bibnamefont {Rosenbluth}}, \bibinfo {author} {\bibfnamefont {M.~N.}\ \bibnamefont {Rosenbluth}}, \bibinfo {author} {\bibfnamefont {A.~H.}\ \bibnamefont {Teller}},\ and\ \bibinfo {author} {\bibfnamefont {E.}~\bibnamefont {Teller}},\ }\href {https://doi.org/10.1063/1.1699114} {\bibfield  {journal} {\bibinfo  {journal} {J. Chem. Phys.}\ }\textbf {\bibinfo {volume} {21}},\ \bibinfo {pages} {1087} (\bibinfo {year} {1953})}\BibitemShut {NoStop}%
\bibitem [{\citenamefont {Cao}\ \emph {et~al.}(2020)\citenamefont {Cao}, \citenamefont {Chen}, \citenamefont {Jin}, \citenamefont {Liu},\ and\ \citenamefont {Li}}]{Cao2020ACSAMI}%
  \BibitemOpen
  \bibfield  {author} {\bibinfo {author} {\bibfnamefont {B.}~\bibnamefont {Cao}}, \bibinfo {author} {\bibfnamefont {S.}~\bibnamefont {Chen}}, \bibinfo {author} {\bibfnamefont {X.}~\bibnamefont {Jin}}, \bibinfo {author} {\bibfnamefont {J.}~\bibnamefont {Liu}},\ and\ \bibinfo {author} {\bibfnamefont {T.}~\bibnamefont {Li}},\ }\href {https://doi.org/10.1021/acsami.0c18483} {\bibfield  {journal} {\bibinfo  {journal} {ACS Applied Materials \& Interfaces}\ }\textbf {\bibinfo {volume} {12}},\ \bibinfo {pages} {57245} (\bibinfo {year} {2020})}\BibitemShut {NoStop}%
\bibitem [{\citenamefont {Jin}\ \emph {et~al.}(2022)\citenamefont {Jin}, \citenamefont {Chen},\ and\ \citenamefont {Li}}]{jin2022coexistence}%
  \BibitemOpen
  \bibfield  {author} {\bibinfo {author} {\bibfnamefont {X.}~\bibnamefont {Jin}}, \bibinfo {author} {\bibfnamefont {S.}~\bibnamefont {Chen}},\ and\ \bibinfo {author} {\bibfnamefont {T.}~\bibnamefont {Li}},\ }\href {https://doi.org/10.1038/s43246-022-00289-5} {\bibfield  {journal} {\bibinfo  {journal} {Communications Materials}\ }\textbf {\bibinfo {volume} {3}},\ \bibinfo {pages} {66} (\bibinfo {year} {2022})}\BibitemShut {NoStop}%
\bibitem [{\citenamefont {Chen}\ \emph {et~al.}(2024)\citenamefont {Chen}, \citenamefont {Jin}, \citenamefont {Zhao},\ and\ \citenamefont {Li}}]{chen2024intricate}%
  \BibitemOpen
  \bibfield  {author} {\bibinfo {author} {\bibfnamefont {S.}~\bibnamefont {Chen}}, \bibinfo {author} {\bibfnamefont {X.}~\bibnamefont {Jin}}, \bibinfo {author} {\bibfnamefont {W.}~\bibnamefont {Zhao}},\ and\ \bibinfo {author} {\bibfnamefont {T.}~\bibnamefont {Li}},\ }\href {https://doi.org/10.1103/PhysRevMaterials.8.043805} {\bibfield  {journal} {\bibinfo  {journal} {Physical Review Materials}\ }\textbf {\bibinfo {volume} {8}},\ \bibinfo {pages} {043805} (\bibinfo {year} {2024})}\BibitemShut {NoStop}%
\bibitem [{\citenamefont {Sheppard}\ \emph {et~al.}(2012)\citenamefont {Sheppard}, \citenamefont {Xiao}, \citenamefont {Chemelewski}, \citenamefont {Johnson},\ and\ \citenamefont {Henkelman}}]{Sheppard_JCP_2012_ssNEB}%
  \BibitemOpen
  \bibfield  {author} {\bibinfo {author} {\bibfnamefont {D.}~\bibnamefont {Sheppard}}, \bibinfo {author} {\bibfnamefont {P.}~\bibnamefont {Xiao}}, \bibinfo {author} {\bibfnamefont {W.}~\bibnamefont {Chemelewski}}, \bibinfo {author} {\bibfnamefont {D.~D.}\ \bibnamefont {Johnson}},\ and\ \bibinfo {author} {\bibfnamefont {G.}~\bibnamefont {Henkelman}},\ }\href {https://doi.org/10.1063/1.3684549} {\bibfield  {journal} {\bibinfo  {journal} {The Journal of Chemical Physics}\ }\textbf {\bibinfo {volume} {136}},\ \bibinfo {pages} {074103} (\bibinfo {year} {2012})}\BibitemShut {NoStop}%
\bibitem [{\citenamefont {Henkelman}\ \emph {et~al.}(2000)\citenamefont {Henkelman}, \citenamefont {Uberuaga},\ and\ \citenamefont {Jónsson}}]{Henkelman_JCP_2000_climbing}%
  \BibitemOpen
  \bibfield  {author} {\bibinfo {author} {\bibfnamefont {G.}~\bibnamefont {Henkelman}}, \bibinfo {author} {\bibfnamefont {B.~P.}\ \bibnamefont {Uberuaga}},\ and\ \bibinfo {author} {\bibfnamefont {H.}~\bibnamefont {Jónsson}},\ }\href {https://doi.org/10.1063/1.1329672} {\bibfield  {journal} {\bibinfo  {journal} {The Journal of Chemical Physics}\ }\textbf {\bibinfo {volume} {113}},\ \bibinfo {pages} {9901} (\bibinfo {year} {2000})}\BibitemShut {NoStop}%
\end{thebibliography}%


\begin{thebibliography}{0}%
\makeatletter
\providecommand \@ifxundefined [1]{%
 \@ifx{#1\undefined}
}%
\providecommand \@ifnum [1]{%
 \ifnum #1\expandafter \@firstoftwo
 \else \expandafter \@secondoftwo
 \fi
}%
\providecommand \@ifx [1]{%
 \ifx #1\expandafter \@firstoftwo
 \else \expandafter \@secondoftwo
 \fi
}%
\providecommand \natexlab [1]{#1}%
\providecommand \enquote  [1]{``#1''}%
\providecommand \bibnamefont  [1]{#1}%
\providecommand \bibfnamefont [1]{#1}%
\providecommand \citenamefont [1]{#1}%
\providecommand \href@noop [0]{\@secondoftwo}%
\providecommand \href [0]{\begingroup \@sanitize@url \@href}%
\providecommand \@href[1]{\@@startlink{#1}\@@href}%
\providecommand \@@href[1]{\endgroup#1\@@endlink}%
\providecommand \@sanitize@url [0]{\catcode `\\12\catcode `\$12\catcode `\&12\catcode `\#12\catcode `\^12\catcode `\_12\catcode `\%12\relax}%
\providecommand \@@startlink[1]{}%
\providecommand \@@endlink[0]{}%
\providecommand \url  [0]{\begingroup\@sanitize@url \@url }%
\providecommand \@url [1]{\endgroup\@href {#1}{\urlprefix }}%
\providecommand \urlprefix  [0]{URL }%
\providecommand \Eprint [0]{\href }%
\providecommand \doibase [0]{https://doi.org/}%
\providecommand \selectlanguage [0]{\@gobble}%
\providecommand \bibinfo  [0]{\@secondoftwo}%
\providecommand \bibfield  [0]{\@secondoftwo}%
\providecommand \translation [1]{[#1]}%
\providecommand \BibitemOpen [0]{}%
\providecommand \bibitemStop [0]{}%
\providecommand \bibitemNoStop [0]{.\EOS\space}%
\providecommand \EOS [0]{\spacefactor3000\relax}%
\providecommand \BibitemShut  [1]{\csname bibitem#1\endcsname}%
\let\auto@bib@innerbib\@empty
\end{thebibliography}%
\onecolumngrid
\newpage
\clearpage


\title{\Large\bfseries Supporting Information of\\[0.3em]
``Anisotropic Short-Range Order Modulates Ferroelectric Switching in Wurtzite ScAlN Alloys''}

\newcommand{\tabincell}[2]{\begin{tabular}{@{}#1@{}}#2\end{tabular}}
\renewcommand{\thefigure}{S\arabic{figure}}
\setcounter{figure}{0}
\renewcommand{\theequation}{S\arabic{equation}}
\setcounter{secnumdepth}{3}

\begin{bibunit}[apsrev4-2]
\maketitle

\onecolumngrid

\section{Computational methods}
All first-principles calculations were performed within density functional theory (DFT) using the projector augmented-wave (PAW) method \cite{Kresse:1999tq,Kresse:1996kg,Kresse:1996vf} as implemented in the Vienna \textit{Ab initio} Simulation Package (VASP) \cite{Kresse:1993ty}. Exchange-correlation effects were treated using the Perdew--Burke--Ernzerhof (PBE) generalized gradient approximation \cite{PBE_GGA_PRL_1996}. Sc$_x$Al$_{1-x}$N alloys were modeled using 108-atom supercells corresponding to a $3\times3\times3$ replication of the wurtzite primitive cell. A plane-wave cutoff energy of 400 eV and a $2\times2\times2$ Monkhorst--Pack $k$-point mesh \cite{Monkhorst:1976cv} were used throughout. Additional calculations using a $3\times3\times2$ $k$-point mesh produced negligible changes in relative energies and switching barriers. Special quasirandom structures (SQS) \cite{Zunger_PRL_1990_SQS} were generated using the Alloy Theoretic Automated Toolkit (ATAT) \cite{Axel_2009_ATAT,VANDEWALLE_2013_mcsqs} at Sc concentrations of 5.6\%, 11.1\%, 22.2\%, 33.3\%, 37.0\%, and 44.4\%. All atomic positions and lattice parameters were fully relaxed using the conjugate-gradient algorithm until the electronic and ionic convergence criteria of $10^{-4}$ eV and $10^{-3}$ eV, respectively, were satisfied.

\section{DFT-based Monte Carlo sampling}

Short-range ordered (SRO) structures were generated using canonical-ensemble Metropolis Monte Carlo (MC) sampling \cite{Metropolis:1953in} combined with first-principles DFT total energies \cite{Cao2020ACSAMI,jin2022coexistence,chen2024intricate}. Trial moves consisted of random exchanges between Sc and Al atoms on the cation sublattice followed by full structural relaxation. The acceptance probability for a trial move from configuration $i$ to configuration $j$ was $P = \min\{1,\exp(-(E_j-E_i)/k_BT)\}$,
where $E_i$ and $E_j$ are the relaxed DFT total energies and $T$ is the sampling temperature.

\begin{figure}[h]
\centering
\includegraphics[width=1.0\linewidth]{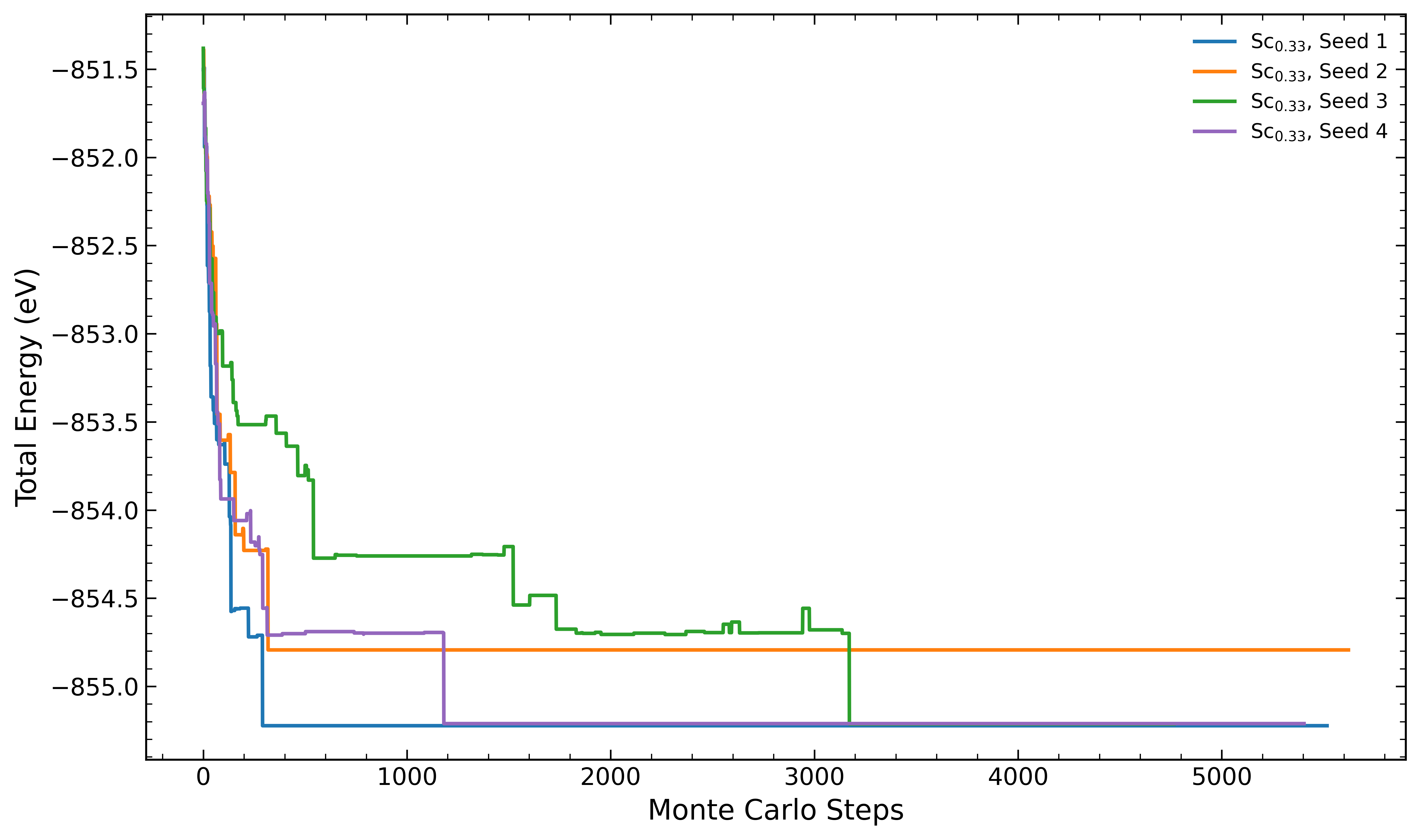}
\caption{Total energy as a function of Monte Carlo (MC) steps for four independent simulations of Sc$_{0.33}$Al$_{0.67}$N starting from different special quasirandom structures (SQSs) generated using different random seeds.
}
\label{fig-mc-Sc0.33}
\end{figure}

\begin{figure}[h]
\centering
\includegraphics[width=1.0\linewidth]{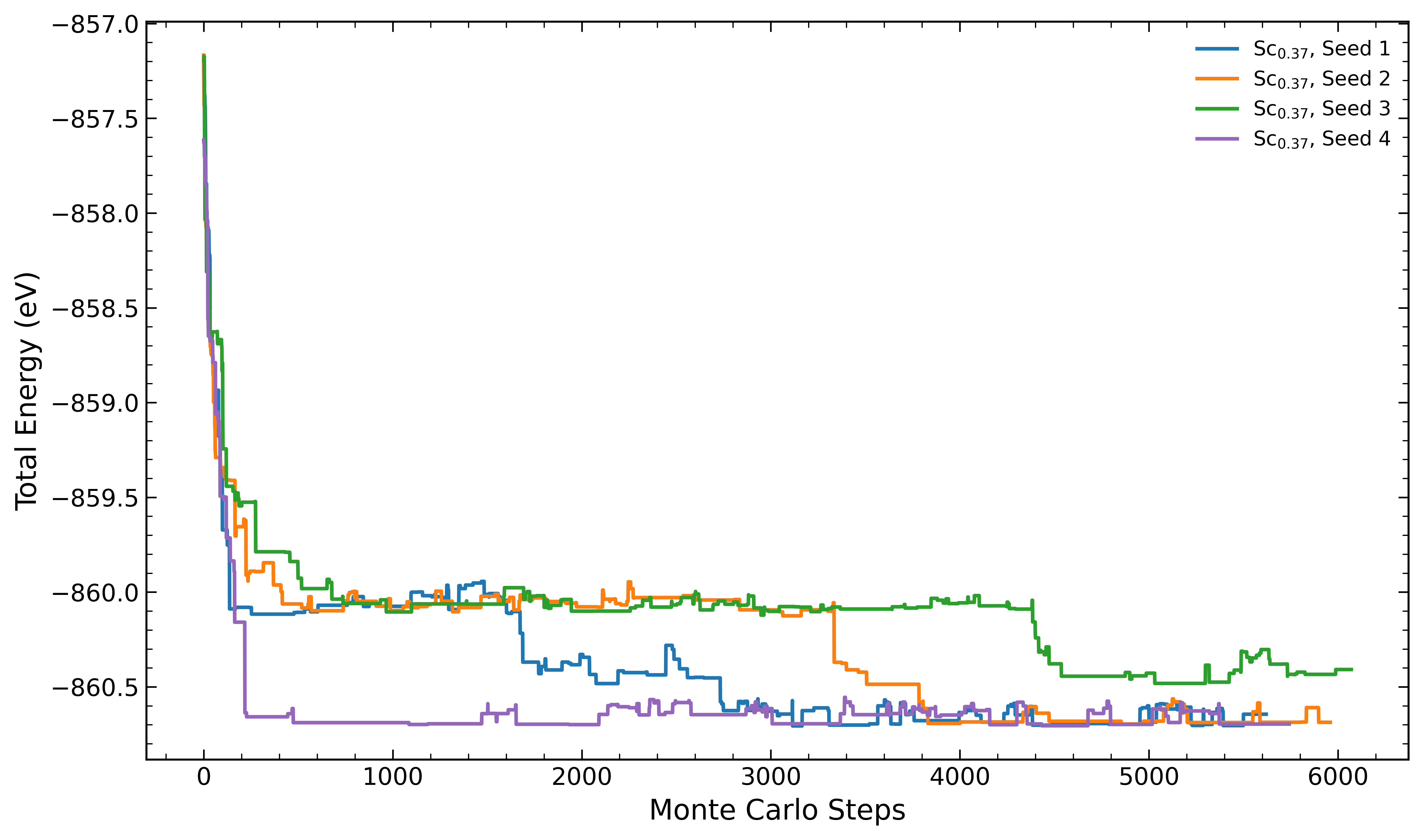}
\caption{Total energy as a function of Monte Carlo (MC) steps for four independent simulations of Sc$_{0.37}$Al$_{0.63}$N starting from different special quasirandom structures (SQSs) generated using different random seeds.
}
\label{fig-mc-Sc0.37}
\end{figure}

For each composition, at least four independent MC trajectories were initialized from distinct SQS configurations generated using different random seeds within the ATAT/mcsqs framework \cite{Axel_2009_ATAT,VANDEWALLE_2013_mcsqs}. MC/DFT sampling was performed at Sc concentrations of 5.6\%, 11.1\%, 22.2\%, 33.3\%, 37.0\%, and 44.4\% to characterize composition-dependent chemical ordering and configurational energetics. For Sc concentrations of 5.6\%, 11.1\%, and 22.2\%, each trajectory was propagated for more than 2000 MC steps at 300 K. For Sc concentrations of 33.3\%, 37.0\%, and 44.4\%, each trajectory was propagated for more than 4000 MC steps at 300 K (see, for example, Fig.~\ref{fig-mc-Sc0.33} and Fig.~\ref{fig-mc-Sc0.37}). Additional MC/DFT simulations were performed for Sc$_{0.37}$Al$_{0.63}$N at 300 K, 500 K, 700 K, and 1600 K to generate configurations with different degrees of local chemical ordering.

\section{Solid-state nudged elastic band calculations}

Ferroelectric switching pathways were calculated using the solid-state nudged elastic band (SS-NEB) method \cite{Sheppard_JCP_2012_ssNEB}. Initial pathways were generated by linear interpolation between fully relaxed polarization states of opposite orientation. Unlike conventional NEB, SS-NEB allows simultaneous relaxation of both lattice vectors and atomic coordinates, enabling determination of minimum-energy switching pathways under zero external stress.

All SS-NEB calculations were performed using the VASP Transition State Theory tools developed by Henkelman \textit{et al.} \cite{Henkelman_JCP_2000_climbing}. The initial and final polar structures were fully relaxed using an electronic convergence criterion of $10^{-6}$ eV, with residual forces below 0.01 eV/\AA\ on every atom. To assess configurational variability, at least four independent SQS structures and four independent SRO structures were analyzed for each Sc composition. Reported averages and standard deviations were computed over these independent configurations. Although some configuration-dependent variation is observed, SRO structures consistently exhibit larger switching barriers than the corresponding SQS structures across the entire composition range investigated.

\section{Convergence of SS-NEB calculations with respect to the number of images}
\begin{figure}[h]
\centering
\includegraphics[width=1.0\linewidth]{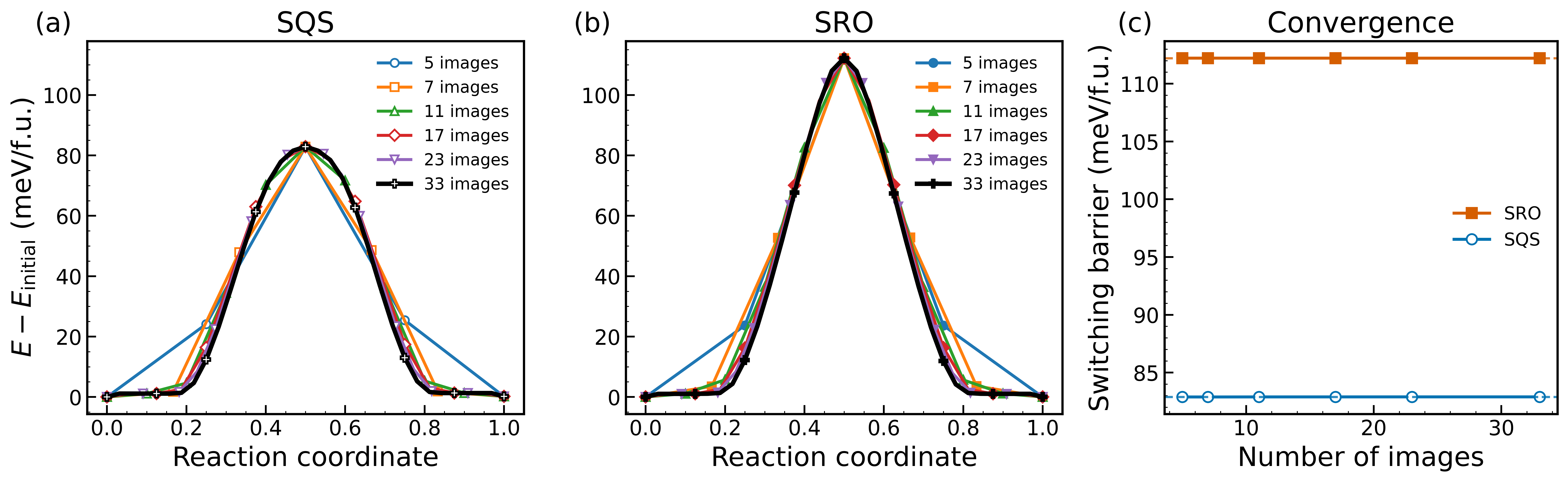}
\caption{Convergence of solid-state nudged elastic band (SS-NEB) calculations for polarization switching in Sc$_{0.33}$Al$_{0.67}$N. (a) Relative energy profiles along the switching pathway for the special quasi-random structure (SQS) model obtained using different numbers of NEB images. (b) Corresponding energy profiles for the short-range ordered (SRO) model. Energies are referenced to the initial state and reported in meV/f.u. The profile obtained using the largest number of images is highlighted in black and serves as the converged reference pathway. (c) Switching barrier as a function of the number of NEB images for the SQS and SRO models. Horizontal dashed lines denote the barrier obtained from the largest-image calculation. 
}
\label{fig-convergence}
\end{figure}

Figure~\ref{fig-convergence} shows the convergence of the calculated switching barrier with respect to the number of SS-NEB images. The switching barrier differs by less than 0.01 meV/f.u. between calculations employing 5 and 33 images, demonstrating excellent convergence with respect to pathway discretization. All SS-NEB calculations reported in this work therefore employed at least 7 images.

\section{Energetic origin of anisotropic short-range order: a controlled motif comparison}

\begin{figure}[h]
\centering
\includegraphics[width=0.92\linewidth]{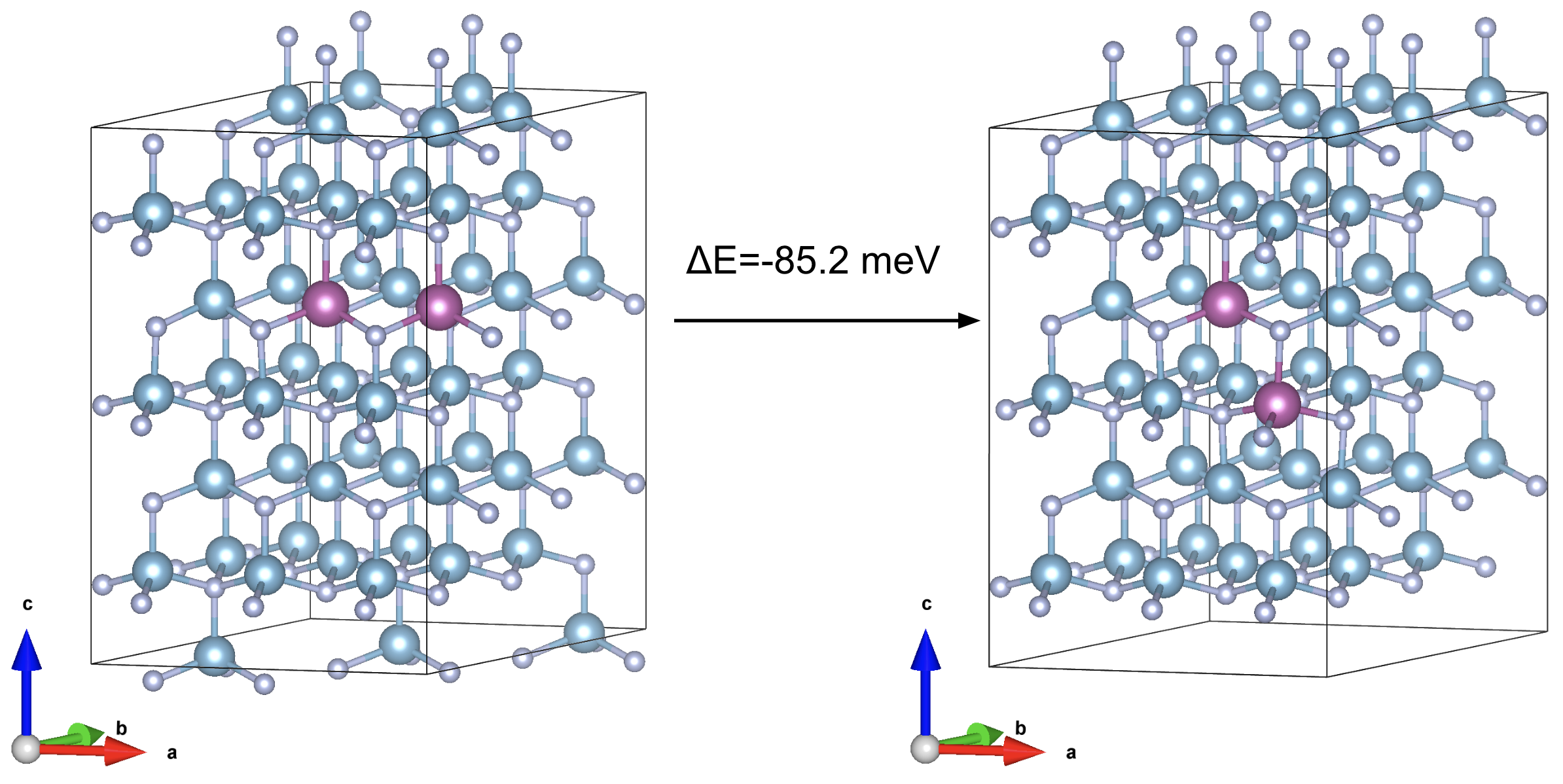}
\caption{
 Two identical 108-atom Sc$_{0.037}$Al$_{0.963}$N supercells differing only in the placement of a single Sc--N--Sc motif. Sc and Al atoms are represented by larger purple and smaller light-green spheres, respectively. After full structural relaxation, the configuration containing a cross-plane Sc--N--Sc motif (right) is lower in energy than that containing an in-plane Sc--N--Sc motif (left) by 85.2 meV. This result indicates that cross-plane Sc--N--Sc motifs are energetically favored over in-plane Sc--N--Sc motifs.
}
\label{fig-motif-cross-Sc-N-Sc}
\end{figure}

\begin{figure}[h]
\centering
\includegraphics[width=0.92\linewidth]{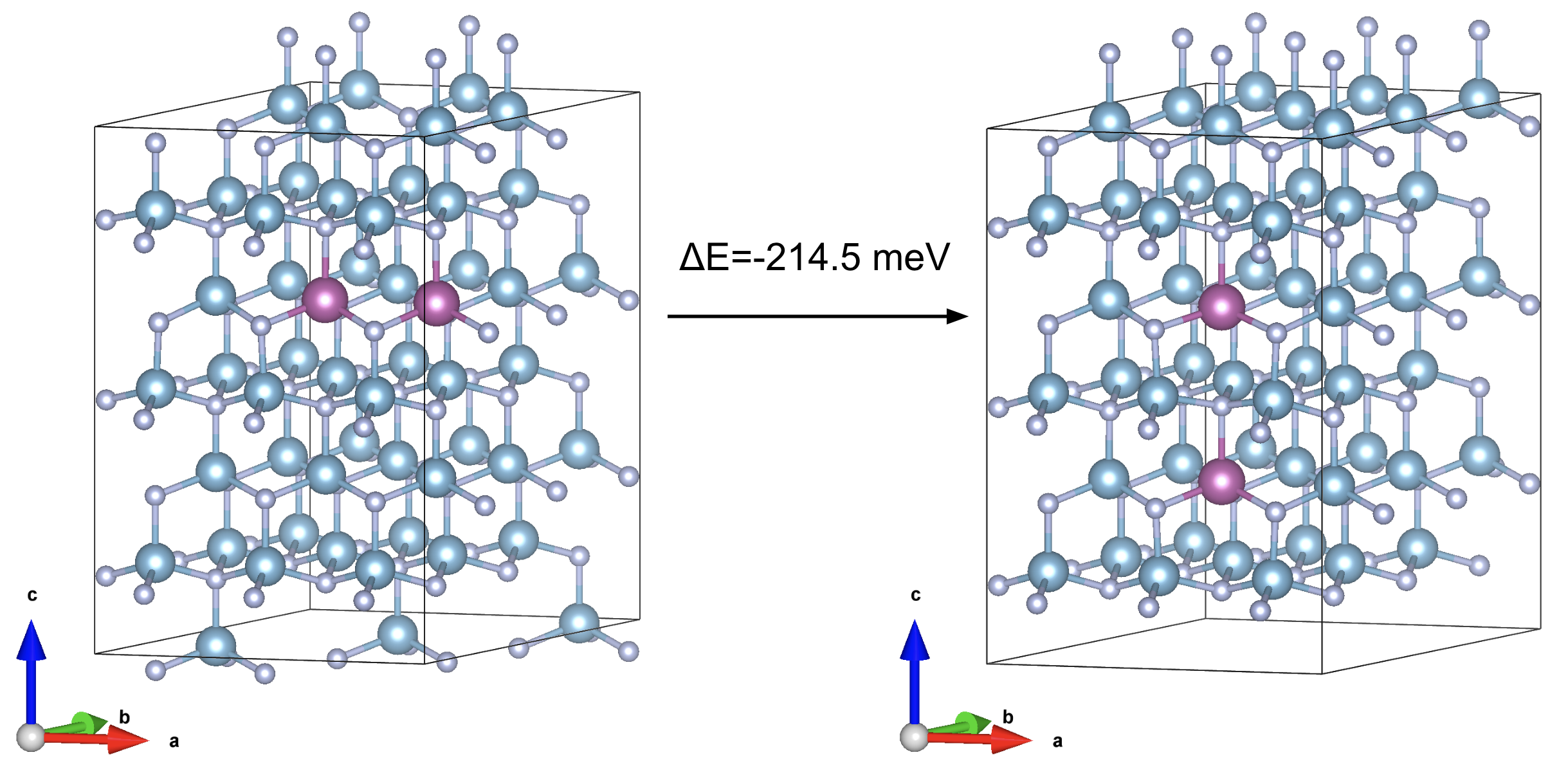}
\caption{
 Two identical 108-atom Sc$_{0.037}$Al$_{0.963}$N supercells differing only in the placement of a single local motif. Sc and Al atoms are represented by larger purple and smaller light-green spheres, respectively. After full structural relaxation, the configuration containing a columnar Sc--N--Al--N--Sc motif (right) is lower in energy than that containing an in-plane Sc--N--Sc motif (left) by 214.5 meV. This result indicates that columnar Sc--N--Al--N--Sc motifs are energetically favored relative to isolated in-plane Sc--N--Sc motifs.
}
\label{fig-motif-columnar-Sc-N-Al-N-Sc}
\end{figure}

To elucidate the energetic origin of the anisotropic short-range order, we constructed three otherwise identical 108-atom Sc$_{0.037}$Al$_{0.963}$N supercells that differ only in the local motif
under investigation. The supercells were identical in composition, cell size, and overall atomic arrangement, enabling direct comparison of motif energetics. After full structural relaxation, a cross-plane Sc--N--Sc motif is lower in energy than an in-plane Sc--N--Sc motif by 85.2 meV (Figure~\ref{fig-motif-cross-Sc-N-Sc}). A columnar Sc--N--Al--N--Sc motif is further stabilized, lowering the total energy by 214.5 meV relative to the in-plane Sc--N--Sc motif (Figure~\ref{fig-motif-columnar-Sc-N-Al-N-Sc}). These calculations establish a clear energetic hierarchy among local motifs, with columnar Sc--N--Al--N--Sc motifs exhibiting the largest stabilization,
thereby providing a microscopic explanation for the anisotropic short-range
order observed in the MC/DFT simulations.

\section{Bond-angle distributions}
\begin{figure}[h]
\centering
\includegraphics[width=1.0\linewidth]{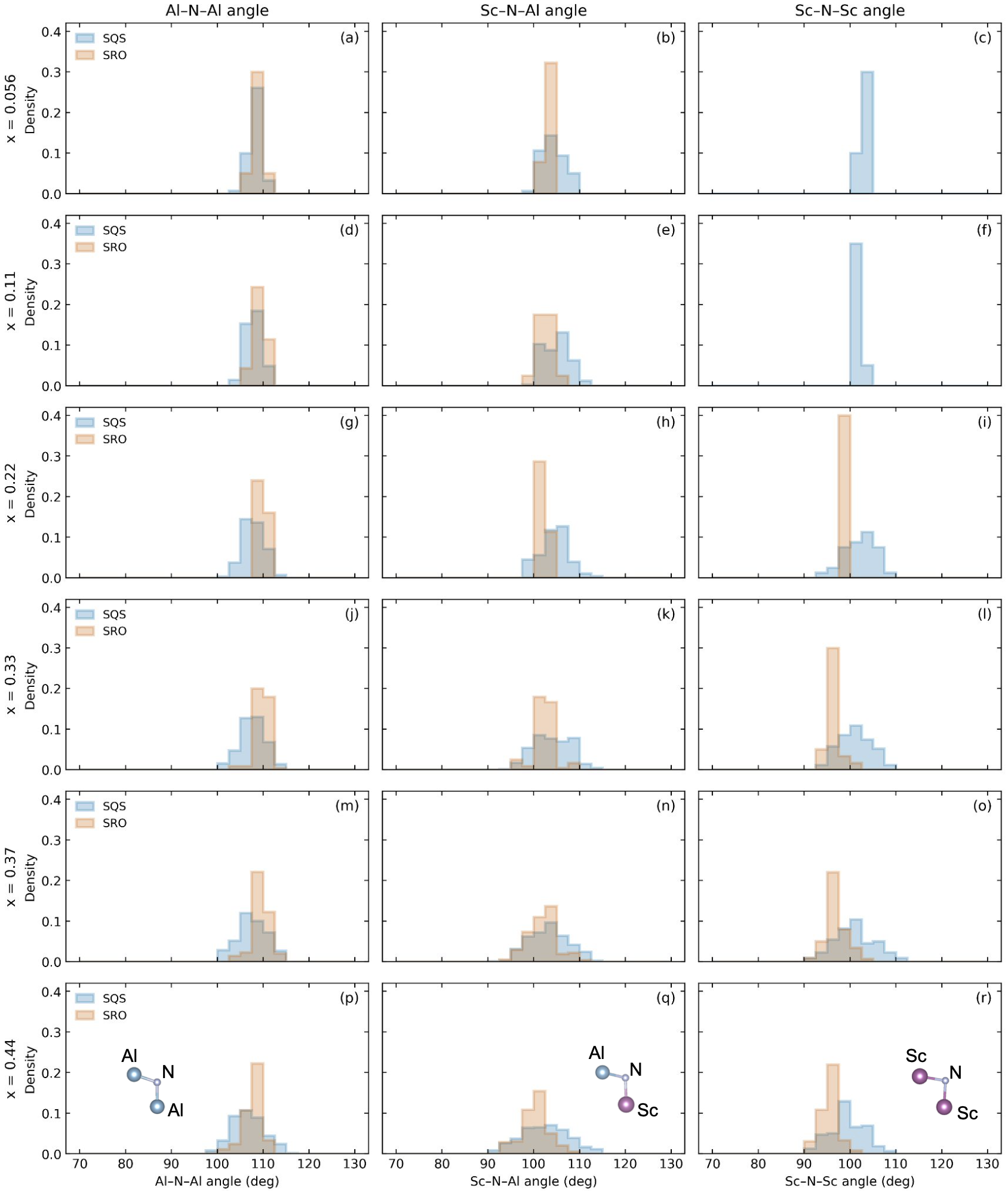}
\caption{
Probability density distributions of cross-plane Al--N--Al, Sc--N--Al, and Sc--N--Sc bond angles (see insets for illustrative graphs) in SQS and SRO Sc$_x$Al$_{1-x}$N structures for $x =$ 0.056, 0.11, 0.22, 0.33, 0.37, and 0.44. Relative to SQS structures, SRO structures exhibit systematically narrower bond-angle distributions, indicating reduced angular disorder in the anisotropic local environments associated with polar connectivity.
}
\label{fig-angle}
\end{figure}

Figure~\ref{fig-angle} shows the distributions of cross-plane metal--N--metal bond angles across multiple Sc compositions. Relative to SQS structures, SRO structures exhibit systematically narrower bond-angle distributions, consistent with the reduced configurational disorder produced by
anisotropic chemical ordering.

\clearpage
\putbib[Refs-SRO]
\end{bibunit}

\end{document}